\documentclass[%
 reprint,
 amsmath,amssymb,
 aps,
 prx,
 superscriptaddress,
 longbibliography,
]{revtex4-1}

\usepackage{graphicx}
\usepackage{dcolumn}
\usepackage{bm}

\begin{document}

\preprint{APS/123-QED}

\title{Leveraging Chaos for Wave-Based Analog Computation: 
\\Demonstration with Indoor Wireless Communication Signals}

\author{Philipp del Hougne}
 \email{philipp.delhougne@gmail.com}
 \affiliation{Institut Langevin, CNRS UMR 7587, ESPCI Paris, PSL Research University, 1 rue Jussieu, 75005 Paris, France}
\author{Geoffroy Lerosey}%
\affiliation{Greenerwave, ESPCI Paris Incubator PC'up, 6 rue Jean Calvin, 75005 Paris, France}%

\date{\today}

\begin{abstract}
In sight of fundamental thermal limits on further substantial performance improvements of modern digital computational processing units, wave-based analog computation is becoming an enticing alternative. A wave, as it propagates through a carefully tailored medium, performs the desired computational operation. Yet, the necessary designs are so intricate that experimental demonstrations will necessitate further technological advances. Here, we show that, counterintuitively, the carefully tailored medium can be replaced with a random medium, subject to an appropriate shaping of the incident wave front. Using tunable metasurface reflect-arrays, we demonstrate our concept experimentally in a chaotic microwave cavity. We conclude that off-the-shelf wireless communication infrastructure in combination with a simple reflect-array suffices to perform analog computation with Wi-Fi waves reverberating in a room. \begin{description}
\item[DOI]
\end{description}
\end{abstract}

\pacs{Valid PACS appear here}
\maketitle


\section{Introduction}\label{Introduction}

\begin{figure*}[t]
	\begin{center}
\includegraphics [width=15cm] {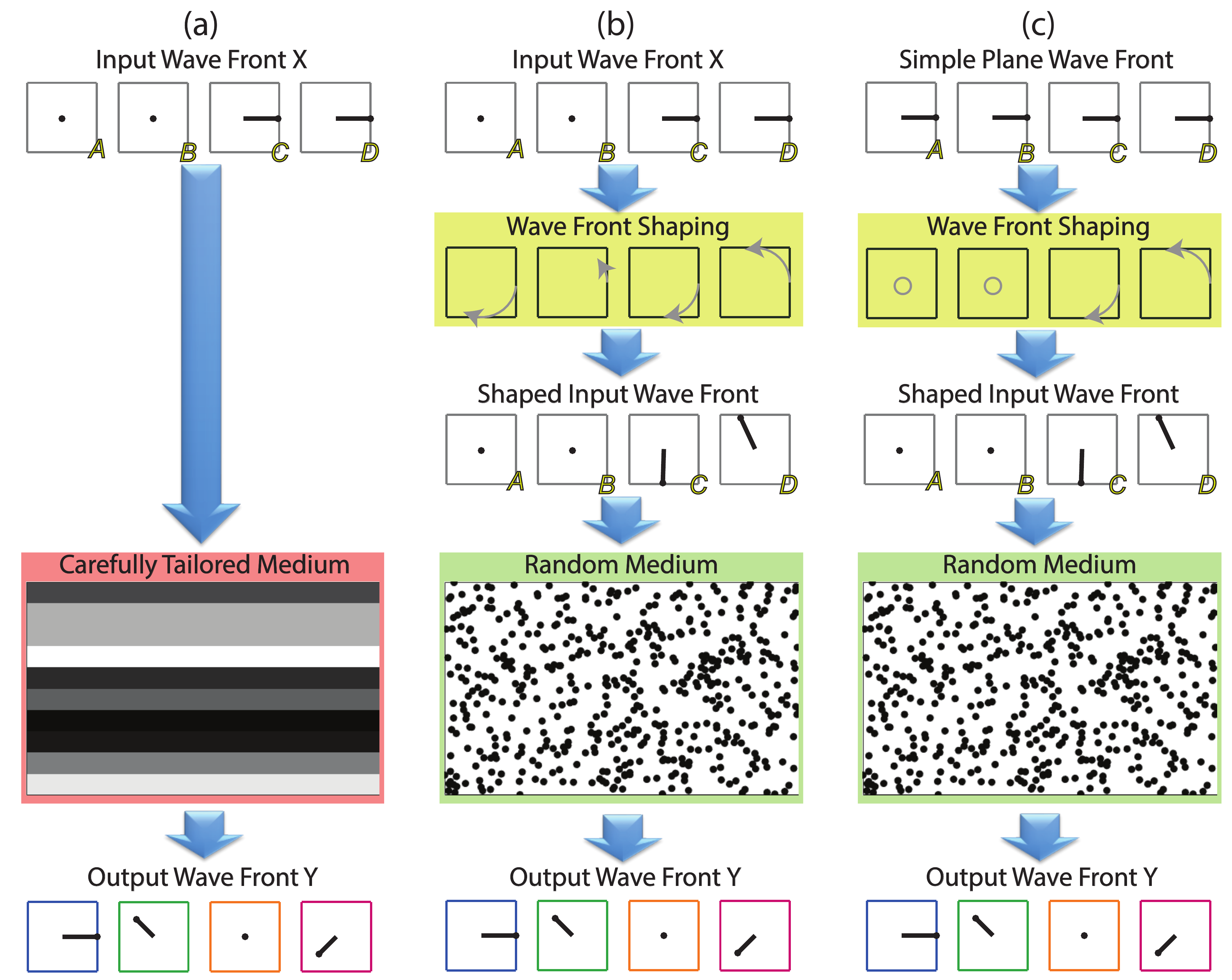}
	\caption{Schematic comparison of different paradigms for wave-based analog computation (WBAC), illustrated for a $4\times4$ discrete Fourier transform operation on input vector $\mathrm{X} = [0 \ 0 \ 1 \ 1]$. Each complex-valued vector entry is visualized as an arrow in the complex plane. (a) The conventional approach performs the desired operation as the input wave front interacts with a medium of carefully tailored scattering properties. (b) We propose to use any random medium instead of a carefully tailored one --- at the cost of appropriately shaping the input wave front before it interacts with the random medium. The shaping of a wave front segment may include modulating its phase, as in this schematic, and/or its amplitude. (c) We further suggest a modified version of (b) in which the wave front shaping step additional encodes the desired input vector $\mathrm{X}$ into the wave front such that any impinging wave front, e.g. a simple plane wave, can be used. The wave front entering the random medium is then the same as in (b).} 
    \label{fig0}
	\end{center}
\end{figure*}

The first steps in the history of computation were analog mechanical devices designed for \textit{specific} tasks such as the Thomson brother's Harmonic Analyzer to determine an input's Fourier representation \cite{RevJalali}. With the dawn of the digital age, electronic \textit{general-purpose} processors took over for good. For a long time, their Moore's Law march of exponential performance improvements made any efforts to design specific-purpose devices redundant \cite{moore}. Nowadays, the underlying approach of squeezing more and more transistors into a unit chip area 
%
is confronted with fundamental thermal limits: already today, only a certain amount of transistors on a processor can be powered at any given time to avoid a silicon meltdown \cite{DarkSilicon}. This so-called \textit{dark silicon} problem recently sparked renewed interest in specific-purpose units conceived to efficiently meet specific computation needs \cite{TrueNorth,GoogleTPU}. One route pursued by all major technology companies are specialized hardware devices based on application-specific integrated circuits or field programmable gate arrays, capable of performing linear operations like matrix-vector multiplications very efficiently. Yet, at some point the same thermal constraints will thwart further performance improvements, too. There is hence a clear motivation to identify fundamentally different computation paradigms. One enticing idea is wave-based analog computation (WBAC) which proposes to swap the electronic circuits for materials with carefully tailored scattering properties such that the computational operation is performed on an incident wave front as it interacts with the material --- see Fig.~\ref{fig0}(a).

Some examples of such materials performing certain operations are well-known free-space optical components: a lens yields the Fourier transform of an impinging wave front \cite{GoodmanBookFourierOptics,reck1994experimental}, but the setup remains very bulky. Recent efforts on identifying appropriate materials for WBAC can be summarized in three groups that bring about different challenges with respect to (re-)configurability and fabrication practicality:

(i) \textit{Leverage a specific physical effect to perform one specific operation enabled by that effect}. A notable and experimentally demonstrated example of this approach used interference phenomena arising from surface plasmon excitation to perform spatial differentiation \cite{FanNatComm}. While this specific operation is certainly of high relevance to high-throughput real-time image processing, it depends on the physics of the leveraged effect and thus inherently does not enable one to configure the device to perform a desired operation.

(ii) \textit{Design a metamaterial block to perform one specific desired operation}. A suitably designed metamaterial block was numerically shown to enable the implementation of one desired transfer function associated with one desired computational operation \cite{EnghetaScience,monticonePRL}. While conceptually appealing due to the significant miniaturization potential, engineering such a complex structure of meta-atoms in practice with sub-micrometer tolerances remains a challenge to be mastered \cite{sihvolaScience}. Even slight fabrication imperfections would inhibit the device from performing the desired operation. Moreover, once fabricated, the device could not be reconfigured to perform a different operation.

(iii) \textit{Program a coherent nanophotonic circuit to perform a desired operation}. Recent advances in silicon photonics enable the fabrication of a mesh of interconntected Mach-Zehnder interferometers, programmable via phase shifters in the waveguide arms to implement arbitrary linear operations \cite{miller2013self,ScienceInterf,ribeiro2016demonstration,MITonn,MillerInterfRev}. The ability to reconfigure the performed matrix-vector multiplication and some first experimental demonstrations reveal the promise held by this emerging research area. 
Precise beamsplitters requiring elaborate fabrication techniques and scalability issues remain critical challenges.

The crux in approaches (ii) and (iii), which enable implementing a desired operation, lies in intricate material designs that are difficult (if not impossible, to date) to fabricate. Here, we propose to overcome any material fabrication and (re-)configuration constraints by using a medium with random instead of carefully tailored scattering properties --- at the much lower price of appropriately shaping the input wave front before it interacts with the random medium. In principle, our concept preserves all the essential advantages brought about by WBAC, in particular enhanced computation speed and power efficiency, and proposes a way of enjoying them without intricate fabrication procedures. We \textit{experimentally} demonstrate the validity of our concept in the microwave domain using a chaotic cavity as random medium and a simple phase-binary metasurface reflect-array to shape the wave field. 
The fact that we only add a metasurface to a setup emulating standard indoor wireless communication 
infrastructure shows the ease of implementing our concept in practice, in sharp contrast to any of the current aforementioned WBAC schemes. 
Then, exploiting the low time cost of WBAC, we demonstrate a scaled-up version under time-sequential operation.
Finally, we discuss perspectives for practical implementations and how they may compete with traditional electronic processors in terms of speed and energy efficiency.

\section{Operation Principle}\label{OperationPrinciple}

In the present work, we focus on implementing linear matrix-vector multiplications. As evidenced by substantial industrial efforts to design electronic hardware accelerators (ASICs, FPGAs) specialized on such operations, they are a key ingredient of many computational needs. Artificial neural networks capable of deep learning rely on ``layers of neurons'', each of which consists of a linear matrix-vector multiplication and a non-linear activation function \cite{DeepLearning}. The heavy reliance on linear transformations is the computational bottleneck  with standard electronic processors. More generally, many key operations in physics and engineering can be expressed as linear operations, such as the discrete Fourier transform that we will use as example in the following. Despite our focus on linear operations, our proposed approach could in principle be extended to performing non-linear operations, for instance by using non-linear wave front shaping devices \cite{NL_Sievenpiper}.

At first sight, the complete scrambling of a wave front as it propagates through a complex medium may seem contrarious to our objective of performing computation for which a very specific operation is to be carried out on the impinging wave front \cite{freund}. Indeed, for a long time the random secondary sources (scatterers or reflectors) constituting a complex medium were considered undesirable and detrimental.  
Various novel techniques, notably time reversal and wave front shaping (WFS), broke that paradigm by leveraging the secondary sources as degrees of freedom \cite{TR_fink,mosk_SLM}.
Inspired by aberration corrections with deformable mirrors in astronomy, WFS applies the appropriate phase and/or amplitude modulation to each segment of an impinging wave front such that after propagation through the complex medium the desired output wave front is obtained \cite{mosk_SLM,NatPhotReview,RotterGigan}. Initially conceived to counteract the scrambling of a wave front in space and/or time \cite{mosk_SLM,popoff_prl,aulbach_STF,katz_STF,publikation3}, WFS later-on paved the path towards harnessing a medium's complexity, as in the case of focusing beyond the Rayleigh limit achievable in homogeneous media \cite{mosk_disorder4perfectFOC,choi2011subwavelengthFOC,park2013subwavelength}. 
Notably, WFS in complex media enabled the demonstration of programmable beam-splitters, the investigation of complex quantum-walks as well as custom-tailored mode sorting \cite{pinske,hugo,fickler}.
Moreover, random projections through multiply scattering media were recently shown to be a potentially powerful data pre-processor, for instance to approximate kernels in machine learning \cite{kernels}.

\begin{figure}[t]
	\begin{center}
\includegraphics [width=7cm] {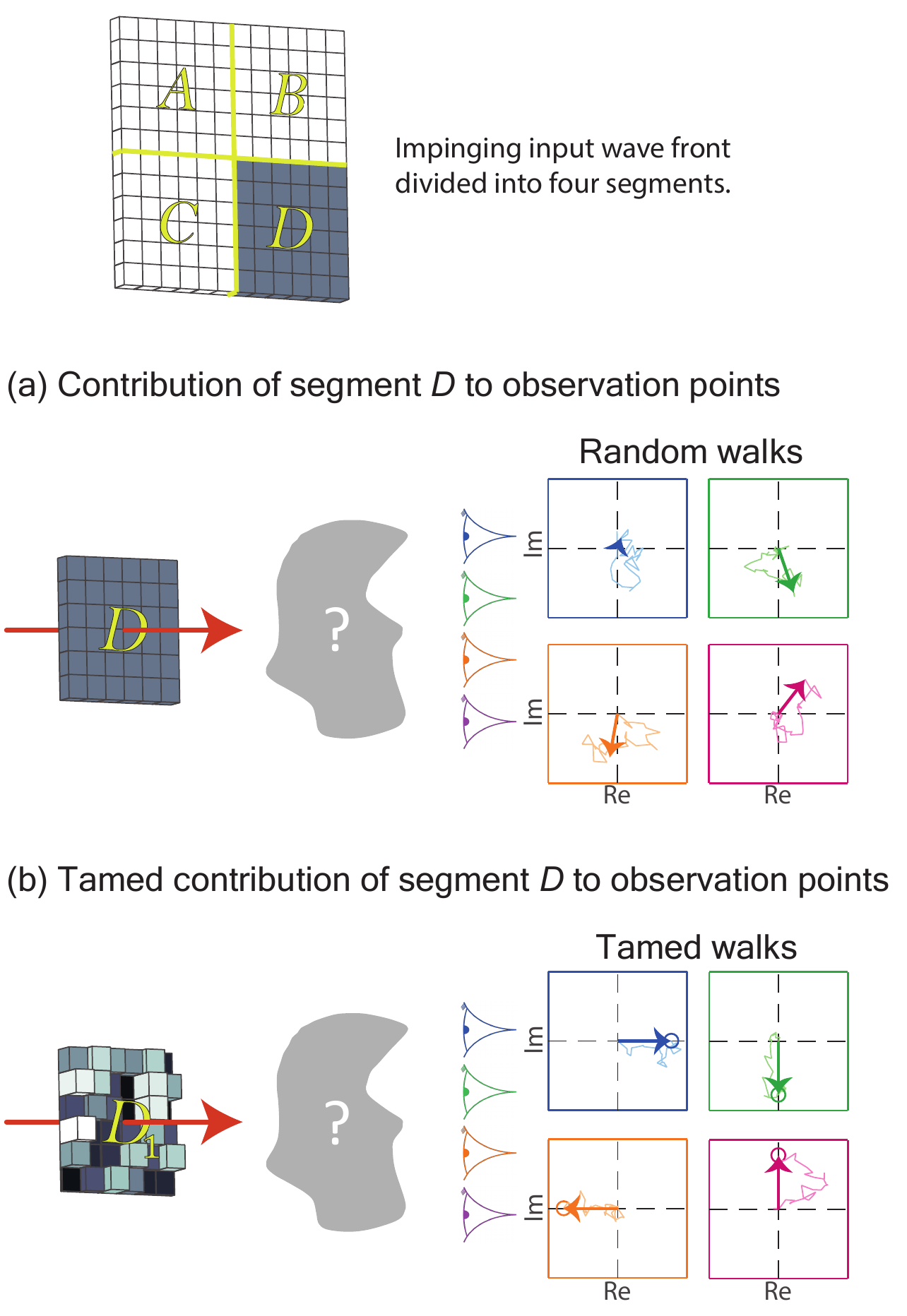}
	\caption{The impinging input wave front in Fig.~\ref{fig0}(b) consists of four segments taking the role of a four-element input vector. This figure illustrates the contribution of one of these segments, $D$, to all four observation points after propagation through the complex medium, contrasting instances without, (a), and with, (b), wave front shaping of the impinging wave front. Each segment consists of $n_{\mathrm{seg}}$ pixels, corresponding to a discretized wave front shaping device, with $n_{\mathrm{seg}} \geq 1$ in general.     (a) A plane wave entering a complex medium gets completely scrambled.  
    At each observation point, stringing together each pixel's contribution yields a random walk whose resultant (the collective contribution from segment $D$) is indicated by a bold arrow.     
    (b) The discretized wave front of segment $D$ can be shaped such that after propagation through the complex medium the resultants at each observation point mimic the  entries of the corresponding row of $\mathbf{G}$. The latter are indicated by circles; for our example of a discrete Fourier transform the forth row of $\mathbf{G}$ reads $ [1\ -i\ -1\ i]$.} 
	\label{fig_principle}
	\end{center}
\end{figure}

A linear computation operation may be expressed as a matrix $\mathbf{G}$ that is applied to an input vector $\mathrm{X}$, yielding the output vector $\mathrm{Y}=\mathbf{G}\mathrm{X}$. To illustrate our general protocol, in Fig.~\ref{fig0}(b) 
we consider the case of a $4\times4$ discrete Fourier transform operation for concreteness. We divide the incident wave front into four segments, $A$ to $D$, that will take the role of each of the four entries of the input vector, in our example $\mathrm{X}=[0\ 0\ 1\ 1]$. Each segment has the amplitude and phase of the corresponding input vector entry. Similarly, we observe the wave field after propagation through the medium at four independent points (color-coded) that will take the role of the output vector's entries. 
Without any shaping, each wave front segment will yield different random contributions to each of the four observation points.
Depending on the discretization of the device used for WFS, each of these four wave front segments may in fact cover several pixels of the discretized WFS device, as illustrated in Fig.~\ref{fig_principle}. Then, as shown in Fig.~\ref{fig_principle}(a), each of the pixels will have its own random contribution to each of the output points, and the overall contribution of segment $D$ is obtained by stringing together the phasors from the contributions of all pixels of which segment $D$ consists. This resembles a random walk in the complex Fresnel representation.

In a one-off initiation step, we first characterize the impact of each pixel of the WFS device on the output wave front, yielding its ``impact matrix'' (IM) $\mathbf{H}$. The entry $H_{i,j}$ gives the gain between the input pixel indexed $i$ and the output observation point indexed $j$. The nature of the IM depends on the system under consideration; in the simplest case of a transmission geometry, as in the schematic in Fig.~\ref{fig0} and standard optical WFS experiments, the IM is simply the system's transmission matrix as in Ref.~\cite{popoff_prl}. In any case does the IM provide an open-loop characterization of the random medium, enabling one to identify numerically specific configurations of the WFS device that satisfy specific objectives. In particular, by shaping the wave front segment, the random walks seen in Fig.~\ref{fig_principle}(a) can be converted into tamed walks whose resultants mimic the corresponding entries of $\mathbf{G}$, as shown in Fig.~\ref{fig_principle}(b) for segment $D$. In other words, WFS enables one to force the contribution from each input entry to each output entry to be exactly the one required by the desired operation $\mathbf{G}$. 
Notice the key role played by the random medium by enabling each input segment to contribute to each output point --- without the random medium only diagonal matrices $\mathbf{G}$ could be implemented. 
The shaped wave front for segment $D$ has to simultaneously yield  contributions 
with specific phases and amplitudes to 
all four observation points, a constraint that is much harder to satisfy than simple focusing which only requires aligning the phasors in an arbitrary angular direction at a single observation point. 
Moreover, note that we do not tame the entire field at the observation points, but the contributions from specific wave front segments to the field at the observation points.
Identifying the appropriate configurations for each segment of the WFS device based on the measured IM is the second step of the initiation procedure, and it is performed purely numerically. A key aspect about this initiation procedure is that to perform a different operation $\mathbf{G}$ one does not require any further measurements, one simply goes through the numerical optimization once again. 

Once a library of optimized configurations for each wave front segment is identified, an impinging wave front can be shaped such that its subsequent propagation through the complex medium carries out the desired computational task at the speed of wave propagation, as summarized in Fig.~\ref{fig0}(b). This concept shifts the burden from designing and fabricating the physical layer's scattering properties to identifying and imposing appropriate wave fronts. 
This is substantially simpler both in terms of the optimization problem as well as the experimental implementation. 
Instead of encoding the input vector $\mathrm{X}$ into the impinging wave front, alternatively $\mathrm{X}$ may be encoded directly into the configuration of the WFS device such that a plane impinging wave front can be used. This idea is visualized in Fig.~\ref{fig0}(c). The underlying thought is that most likely in order to have the input vector encoded into the impinging wave front, some WFS device outside the considered WBAC system was employed. Then, however, instead of using essentially two WFS devices directly one after the other, one could use a single WFS device to do both tasks. In Fig.~\ref{fig0}(c) we show how the WFS step simultaneously encodes the input into the impinging plane wave front and adds the modulation required given the random medium to perform the desired operation. Then, the wave front entering the random medium is the same as in Fig.~\ref{fig0}(b).

\section{Demonstration with indoor wireless communication signals}\label{Demonstration}

Having outlined the general principle of operation, we now proceed with an experimental demonstration that seeks to stress (i) the ease of implementation in contrast to hitherto existing WBAC proposals, and (ii) the wide applicability to all types of waves and random media. The first complex medium that comes to mind is usually a multiply scattering one. In the microwave domain, examples thereof exist, such as cities or forests, but they are highly unpractical for experiments. Instead, cavities of irregular geometry lend themselves to practical, well-controlled and stable setups, and are moreover very commonplace. 
Indeed, indoor environments constitute cavities of low quality factor for the microwaves used in wireless communication protocols such as Wi-Fi. Their geometries are in general rather irregular, such that these cavities may be labeled ``random'' or ``chaotic'' from the wave's perspective. The reflections off the irregular cavity walls interfere and give rise to a speckle-like wave field \cite{MIMO_PhysToday}, very much like scattering events in multiply scattering optical media such as thin paint layers or biological tissue \cite{mosk_SLM,choi2015WSforBioMed}. Furthermore, complex microwave cavities are leveraged in fundamental research on quantum chaos \cite{BookStockmann} as well as in applications ranging from security screening \cite{JonahSciRep,TimDMA} and biomedical imaging \cite{winnipeg1}, via sensing \cite{MotionDetector,Localization} and wireless power transfer \cite{Anlage_NLlossyTR,WTP_Smith,harvesting_arXiv,vinay} to electromagnetic compatibility tests 
\cite{hill_electromagnetic_2009}.

\subsection{Shaping Cavity Green's Functions}

Despite the conceptual similarity between multiply scattering media and chaotic cavities, there are important nuances that we will discuss here to motivate our subsequent course of action. 
A convenient way of probing the field in a cavity is to measure the Green's function between two independent positions \cite{stoffregen,SmatrixLegrand,Anlage_S_Matrix,kuhl2005classical}. 
To get a firmer grasp of a chaotic cavity's physics, it is instructive to express the transmitted field between two antennas in terms of the modal contributions. Omitting for clarity both details of the antenna coupling as well as the vector nature of the electromagnetic field, we can write the Green's function $S$ between two points $\mathrm{r_i}$ and $\mathrm{r_j}$ as
\begin{equation}\label{ModalSum}
 S(\mathrm{r_i},\mathrm{r_j},f_0) = \sum_{n=1}^N \frac{\psi_n(\mathrm{r_i})\psi_n(\mathrm{r_j})}{\frac{4\pi^2}{c^2} (f_0^2-f_n^2) + 2 \pi i f_0 \Gamma} ,
\end{equation}
\noindent where $\psi_n$ and $f_n$ are eigenvector and eigenvalue of the $n$th cavity mode that contributes to the Green's function at the working frequency $f_0$. $\Gamma$ is the (average) modal line-width.  An estimate of the number of modes $N$ overlapping at $f_0$, this time accounting for the three-dimensional and polarized nature of the field, can be obtained from Weyl's law \cite{WEYLoriginal,WEYLbook,publikation1}:
\begin{equation}\label{weyl}
 N =  \frac{8\pi  f_0^3}{c^3} \frac{\mathcal{V}}{Q},
\end{equation}
\noindent $\mathcal{V}$ and $Q$ being the cavity's volume and quality factor, respectively. In a chaotic cavity, $\psi_n$ can be modeled as random variable, such that  $S(\mathrm{r_i},\mathrm{r_j},f_0)$ can be interpreted as a random walk of $N$ steps in the complex plane --- just as illustrated in Fig.~\ref{fig_principle}(a).

The crucial remaining question is: How can a cavity Green's function be shaped? The concept of WFS in complex media originally emerged in optics \cite{mosk_SLM} but has recently been transposed to the microwave domain using 
tunable metasurface reflect-arrays, with important applications in telecommunication, imaging, sensing and energy transfer \cite{SMM_PoC,TimDMA,MotionDetector,Localization,winnipeg1,harvesting_arXiv,vinay}. 
Different designs of such reconfigurable metasurfaces have been discussed in the literature \cite{SievenpiperPMC,SievenpiperTunable,SihvolaMetaOverview,SMM_design,TimDMA,SievenpiperReview}; here, we use a simple phase-binary device whose working principle is based on the hybridization of resonances (see Ref.~\cite{SMM_design} and 
Section~A in Ref.~[63] 
for further details). 
We can electronically configure the phase shift of the reflected wave for each of the 
$88$ metasurface elements
to be $0$ or $\pi$. Stated differently, each metasurface element can be programmed to mimic a Dirichlet or a Neumann boundary condition.

By partially altering the cavity's boundary conditions with the tunable metasurface, we have some control over the modal sum in Eq.~\ref{ModalSum} because we essentially modify the cavity eigenmodes $\psi_n$. The degree of control depends both on the cavity's modal overlap which fixes $N$ (see~Eq.~\ref{weyl}) and on the amount of metasurface elements, yielding the capability of ``effectively'' controlling $p$ out of the $N$ modes that contribute to a cavity Green's function at the working frequency \cite{publikation1}. As discussed in Ref.~\cite{publikation1}, the ratio $p/N$ defines distinct regimes: (i) if $N>p$, we control effectively a subset of the contributing modes, such that the random walk can be tamed partially; (ii) if $N\approx p$, we have full control over the modal sum and can tame the walk at wish; (iii) if $N \ll 1 < p$, the cavity spectrum consists of discrete resonances that can be created and shifted at wish. 
 
\subsection{Choosing a convenient regime}

Unlike the simple transmission geometries used in Section~\ref{OperationPrinciple} to illustrate our WBAC scheme, the physics of a chaotic cavity is richer than captured by a simple, linear transmission formalism. Here, the IM captures the impact of each metasurface reflect-array pixel on a selection of Green's function measurements. Since the reverberating wave revisits each pixel of the WFS device multiple times, the impact of a given pixel on the wave field is correlated to some extent to the configuration of the remaining pixels: $\mathbf{H}$ is itself a function of the metasurface reflect-array configuration. A pixel acts not like a single but multiple (secondary) sources, due to reverberation. 

In principle, the realm of machine learning with artificial neural networks \cite{DeepLearning} offers elegant ways of accurately capturing the full IM including the long-range correlations. However, having such a complete forward model would not alter the fact that physically the contributions of a given 
``input'' group of pixels on the WFS device 
to the 
``output'' Green's function measurements 
depend to some extent on how the 
remaining pixels of the WFS device are 
currently configured. 
In other words, in a cavity one cannot hope to be able to shape the contributions of one 
wave front segment independently from the other ones. 
What appears at first sight to be an insurmountable hurdle, can in fact be easily resolved by leveraging the 
cavity
's chaotic nature.

First, we will deliberately measure only a first-order approximation of $\mathbf{H}$ that ignores the long-range correlations. That is, we neglect the dependence of a pixel's wave field impact on the configuration of the remaining pixels, yielding a linear forward model.
Second, we will numerically identify a library of segment configurations that appropriately tame the wave front for the desired computational operation (see Section~\ref{ExperimentalProcedure} for the detailed numerical procedure), supposing our first-order IM is exact. Of course, the computation outcomes that we expect based on our approximated IM, $\mathrm{Y}^{\mathrm{pred}}$, will not be exactly what we will measure, $\mathrm{Y}^{\mathrm{meas}}$, since we did not account for the long-range correlations yet. 
The effect of the latter on the $j$th entry of the computation output can be interpreted as a random phasor $\delta_j$: $Y_j^{\mathrm{meas}} = Y_j^{\mathrm{pred}} + \delta_j$. For a given realization, $\delta_j$ is a deterministic but seemingly arbitrary random complex number that results in an inevitable inaccuracy of our computation outcome $Y_j^{\mathrm{meas}}$. A realization may be defined as one instance of the experiment with some specific (irregular) cavity geometry. A different realization will have the same global parameters (cavity volume, quality factor, \dots) but a different geometry and thus a different random $\delta_j$. Therefore, we can exploit the realization-dependence of $\delta_j$ to average out the intrinsic inaccuracy: $\langle \delta_j \rangle _{\mathrm{realizations}} = 0$, and thus $\langle \mathrm{Y}^{\mathrm{meas}} \rangle _\mathrm{realizations} = \mathrm{Y}^{\mathrm{pred}}$. 
By ensemble averaging over different realizations we are thus able to mitigate the adverse effects of long-range correlations in the IM.

The importance of long-range correlations on the impact of pixel $i$ on the transmission measurement indexed $j$ is defined by two parameters that govern how likely it is that a ray that interacts with pixel $i$ also encounters other pixels during its life-time: (i) the higher $Q$ is, the longer the ray's life-time is and the more pixels other than $i$ it will encounter; (ii) the higher the amount of pixels, the easier it is for the ray to encounter some of the other pixels.
It is instructive to think of an IM entry $H_{i,j}$ as consisting of static contributions from rays that only interacted with pixel $i$ and of variable contributions from rays that interacted also with other pixels of the metasurface reflect-array. 
Ultimately, we average out the effect of the latter such that our approach works as long as the former is non-zero. The impact of a given pixel of the WFS device on the wave field must not completely depend on how the other pixels are configured. 

An ideal fit in sight of these considerations is thus operating in the regime $N>p$ in an electrically large cavity of (relatively) low $Q$. Of course, reverberation is crucial to ensure the cavity acts like a random medium; but limiting $Q$ to the order of $10^2$ rather than $10^3$ or $10^4$ addresses (i). Simultaneously, since $N$ will be large under these conditions (see Eq.~\ref{weyl}), one can fit a large metasurface with many pixels into the cavity and still respect (ii). One might speculate that as $N$ becomes larger than $p$, the metasurface completely dominates the wavefield and the static cavity contribution to $H_{i,j}$ vanishes --- ultimately, confirming this remains a topic for future work. It is clear, however, that the limiting case of very-high-$Q$, small-volume cavities in the regime $N\ll 1<p$ with discrete resonances is inappropriate for our scheme \cite{publikation1,anderson,genack_transport_random_media}. 

\subsection{Experimental Setup}

\begin{figure}[t]
	\begin{center}
\includegraphics [width=\columnwidth] {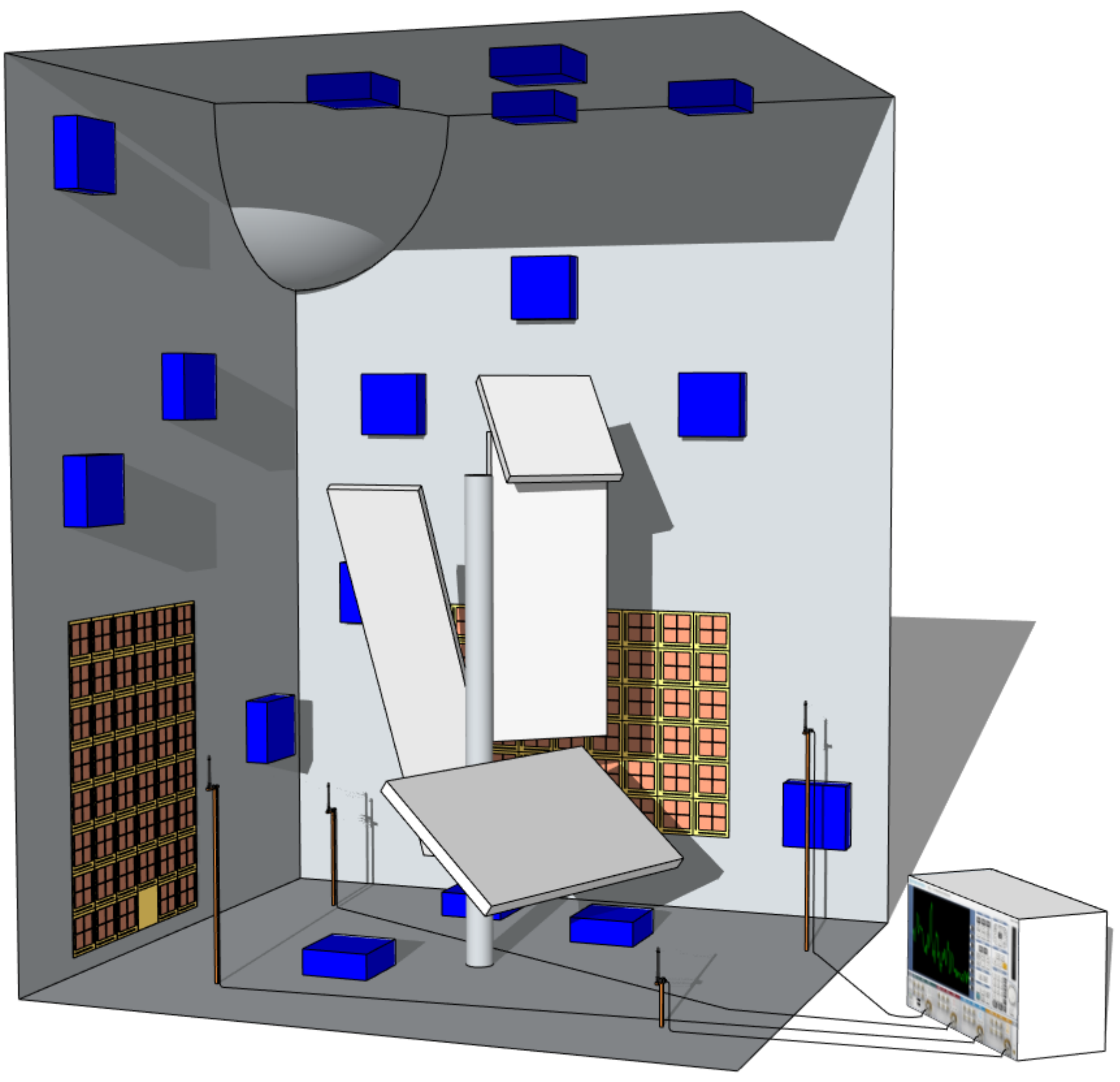}
	\caption{Laboratory setup to test feasibility of analog computation with Wi-Fi waves in an indoor room. About $6\%$ of a chaotic metallic cavity's surface are covered with a metasurface reflect-array consisting of $88$ pixels with tunable reflection coefficient ($r=\pm1$); see Ref.~\cite{SMM_design} and Section~A in Ref.~[63] for the detailed working principle. Electromagnetic absorbers placed on the walls lower the cavity quality factor to $Q=179$, comparable to an indoor room. With a vector network analyzer, we measure the transmission between four simple monopole Wi-Fi antennas, mimicking channel state information between an access point and four mobile devices. A mode-stirrer, rotated in steps of $12^\circ$, enables ensemble averaging over realizations of the experiment under different (irregular) cavity geometries.} 
	\label{fig_setup}
	\end{center}
\end{figure}

In sight of the above considerations, our experimental setup thus consists of an irregularly shaped cavity with (relatively) low quality factor and the ability to access different realizations for ensemble averaging. 
In order to operate under stable, well-controlled conditions, we work in a metallic cavity of irregular geometry ($\mathcal{V}=1.01\times0.86\times1.28\ \mathrm{m}^3 = 1.1 \ \mathrm{m}^3$). By placing absorbers on its walls, we limit to some extent the reverberation with the previously outlined motivation; the resulting quality factor $Q=179$ is comparable to the one we measured in an office room. Given our working frequency $f_0 = 2.65\ \mathrm{GHz}$, there are thus $N\approx 106$ overlapping modes. 
Our metasurface consisting of $n=88$ elements can effectively control about $p\approx 2\times(n/3) = 59$ of these modes (see Section~A in Ref.~[63]
). 
We are thus operating in the first regime where $N>p$, under conditions that are typical for an indoor environment. A schematic is shown in Fig.~\ref{fig_setup}. 
Fig.~S1 in Ref.~\footnote{See Supplemental Material} illustrates with experimental data that in our setup $H_{i,j}$ is indeed dominated by the static cavity contribution, as desired.

We need as many independent single-frequency transmission measurements as our output vector $\mathrm{Y}$ is supposed to have entries. 
We choose to place antennas at arbitrary positions inside the cavity that are at least half a wavelength apart, which is an obvious option given the wave field's speckle-like nature with a correlation length on the order of half a wavelength; alternatively, one may also use a single pair of antennas and multiple working frequencies that are sufficiently different \cite{derodePRE2}. 
In the realistic scenario of indoor wireless communication systems, the channel state information (CSI) between an access point and a wireless device is exactly what we need: a complex-valued transmission measurement between two points inside the cavity. Modern multi-user telecommunication networks routinely sound channels, for instance using beacon signals, for downlink beamforming \cite{CSIbeamforming}. CSI can be accessed with off-the-shelf Intel 5300 cards, and has successfully been extracted and used for instance as fingerprint for object localization \cite{souviksen}. Green's function measurements for different pairs of $\mathrm{r_i}$ and $\mathrm{r_j}$ are thus readily available in commonplace wireless communication infrastructure. In our laboratory implementation, we mimic CSI between wireless devices and a base station by transmission measurements between independent monopole antennas with a vector network analyzer.

Ensemble averaging is easily achieved in our experiment since we can access a new realization either by rotating a mode-stirrer by $12^\circ$ (see Fig.~\ref{fig_setup}) or by changing the operating frequency by more than a correlation frequency (see Section~A in Ref.~[63]
). Both concepts could of course be used in an analogous manner with Wi-Fi in an indoor room. 
The constant motion of inhabitants naturally provides new realizations 
by physically changing the cavity's irregular geometry, analogous to the mode-stirrer in our experiment. 
A single realization would have to be carried out within the coherence time of the medium, which is within reach in sight of 
real-time switching of the PIN diodes controlling the reflect-array with improved electronics \cite{FPGA_metasurface}. 
Alternatively, using the CSI between different pairs of access point and mobile device (e.g. if there are many mobile devices in a room) is another way to 
obtain independent realizations of the experiment without relying on motion inside the room.

\subsection{Experimental Procedure}\label{ExperimentalProcedure}

\textit{Measuring the impact matrix (IM).} --- The entry $H_{i,j}$ quantifies the impact of the reflect-array element indexed $i$ on the transmission between the pair of antennas indexed $j$. We measure the IM using the Hadamard rather than the canonical basis, following the procedure used in Ref.~\cite{popoff_prl}. The elements of the Hadamard basis being $+1$ or $-1$, this is a perfect match to our experiment where we can access exactly these two values for each pixel's reflection coefficient. Moreover, working in the Hadamard basis has a favorable averaging effect as illustrated in the Fig.~S1 in Ref.~[63] and benefits furthermore from an improved signal-to-noise ratio. 

One important difference to Ref.~\cite{popoff_prl}, however, is that we only control a portion of the wave field. As elaborated in the previous section, we operate in the first regime where $N>p$ and we effectively only control about $p/N \approx 56\%$ of the cavity modes. To 
successfully define an IM we subtract from each transmission measurement $Y_j$ the static cavity contribution $U_j$ that cannot be altered by the metasurface. This static contribution corresponds to the uncontrolled cavity modes which hence resist averaging over random reflect-array configurations. We thus obtain $U_j$ as $\langle Y_j \rangle _{\mathrm{random} \ \mathrm{V}}$, where $\mathrm{V}$ is a vector defining the reflect-array configuration, and then we work with $\mathrm{Y}'=\mathbf{H}\mathrm{V}$, where $\mathrm{Y}'=\mathrm{Y}-\mathrm{U}$ as in Ref.~\cite{TM_RevMed}.
Note that this subtraction is only necessary for a cavity implementation of our proposed WBAC scheme; in transmission geometries as in Fig.~\ref{fig0} there is typically no uncontrolled contribution to the output wave front in any case, since the entire incident wave front interacts with the WFS device.

\textit{Numerical identification of wave front segment configurations.} --- Given the IM, we numerically identify the reflect-array configurations for each wave front segment that yield the desired tamed walks illustrated in Fig.~\ref{fig_principle}(b). Consider for concreteness once again the $4\times4$ example given in Figs.~\ref{fig0}~and~\ref{fig_principle}. Let $\mathrm{V}_D$ be the part of vector $\mathrm{V}$ that belongs to segment $D$, let $Y_j^D$ denote the contribution of segment $D$ to the measurement $Y_j$ at observation point $j$ (that is, $Y_j^D$ is a single complex-valued number), and let $\mathrm{H}_{D,j}$ be the corresponding part of the IM $\mathbf{H}$ that links $\mathrm{V}_D$ to $Y_j^D$. In summary, $Y_j^D = \mathrm{H}_{D,j} \mathrm{V}_D$.
Note that selecting the members of segment $D$ amongst all pixels of the WFS device does not have to be done in a regular manner as in our schematic in Fig.~\ref{fig_principle}. In our experiment, we make the selection randomly; a careful selection could in fact be a further parameter of the optimization. 
The objective is to find a $\mathrm{V}_D$, given $\mathrm{H}_{D,1}$ to $\mathrm{H}_{D,4}$, such that \textit{simultaneously} specific values of $Y_1^D$ to $Y_4^D$ are obtained. This can be expressed as a single stacked equation, $[\mathrm{H}_{D,1},\mathrm{H}_{D,2},\mathrm{H}_{D,3},\mathrm{H}_{D,4}] \mathrm{V}_D = [Y_1^D,Y_2^D,Y_3^D,Y_4^D]$. 
In the case of perfect phase and amplitude control over each pixel of the WFS device, the equation is easily solved by calculating the pseudo-inverse of $[\mathrm{H}_{D,1},\mathrm{H}_{D,2},\mathrm{H}_{D,3},\mathrm{H}_{D,4}]$ --- provided each wave front segment is assigned a sufficient number of pixels of the WFS device. 
Indeed, from basic considerations about the singular value decomposition of this stacked matrix it becomes clear that we need to assign at least $\mathcal{N}$ idealistic pixels to each wave front segment 
in order to be able to calculate a meaningful pseudo-inverse 
\footnote{Interestingly, a similar scaling law proportional to $\mathcal{N}^2$ is found for the number of components of a photonic circuit consisting of beam-splitters to implement a desired $\mathcal{N}\times\mathcal{N}$ linear operation \cite{reck1994experimental,MITonn}.}.
In our experimental setup, however, the entries of $\mathrm{V}$ may only take the values of $+1$ or $-1$ due to our phase-binary only metasurface control over the wave front. 
A plethora of methods exist to solve the ill-posed problem resulting from this constraint \cite{LD_binDMD}. 
Here we opt for a home-made solver inspired by standard iterative sequential optimization algorithms used in optical WFS \cite{moskWSalgo}. 

First, we define a normalization factor $\gamma$ between the entries of $\mathbf{G}$ and the target amplitudes of the tamed walks. We choose $\gamma = n_{\mathrm{seg}}^{1/2}\langle |H_{i,j}|\rangle _{i,j}$, where $n_{\mathrm{seg}}$ is the number of pixels in a given wave front segment. This choice corresponds to the expected value of the resultant of a random walk of $n_{\mathrm{seg}}$ steps of step size $\langle |H_{i,j}|\rangle _{i,j}$. 
Next, we define a cost function as $CF = \langle|\gamma G_{D,j} - \mathrm{H}_{D,j} \mathrm{V}_D|\rangle_j$. $CF$ is thus the distance in the complex plane between an entry of $\gamma\mathbf{G}$, say the entry linking input segment $D$ to observation point $j$, 
$\gamma G_{D,j}$, and the corresponding value of $\mathrm{H}_{D,j} \mathrm{V}_D$, averaged over all output points $j$. Starting with a random configuration of $\mathrm{V}_D$, we test element by element if flipping its state reduces $CF$ in which case we update $\mathrm{V}_D$ accordingly. We run 20 loops over all elements. Finally, we repeat the procedure with 250 different random initial $\mathrm{V}_D$ and select the overall best final $\mathrm{V}_D$ yielding the lowest $CF$ as entry for the library. Note that this numerical procedure has not been heavily optimized and more efficient schemes could certainly be found \cite{SimulAnnneal,SciAdvANNinvDesing}. In any case, whatever the numerical cost to identify the wave front segment configurations, this is a one-off effort during the initiation phase of our WBAC scheme.

Given the peculiarity of our chosen cavity implementation in which the WFS device is essentially inside the propagation medium and its pixels are \textit{secondary} sources, it is convenient and most meaningful to adopt the WBAC scheme from Fig.~\ref{fig0}(c) in which the WFS device additionally encodes the input vector $\mathrm{X}$. In principle, additionally encoding the input is a trivial step. However, in our experiment, since we do not have perfect phase and amplitude control on the reflect-array, this is not straightforward. To encode an input vector entry $\alpha$ other than unity into, for example $\mathrm{V}_D$, we thus identify a new configuration for $\mathrm{V}_D$ such that its contribution $\mathrm{H}_{D,j} \mathrm{V}_D$ to the observation point indexed $j$ is $\alpha \gamma G_{D,j}$, simultaneously for all $j$. In this Section, we restrict ourselves to $\alpha \in \{ 0,1\}$ which corresponds to a 1-bit input resolution. We will consider a larger input size and resolution in Section~\ref{scalability}.

\subsection{Experimental Results}

\begin{figure}
	\begin{center}
\includegraphics [width=\columnwidth] {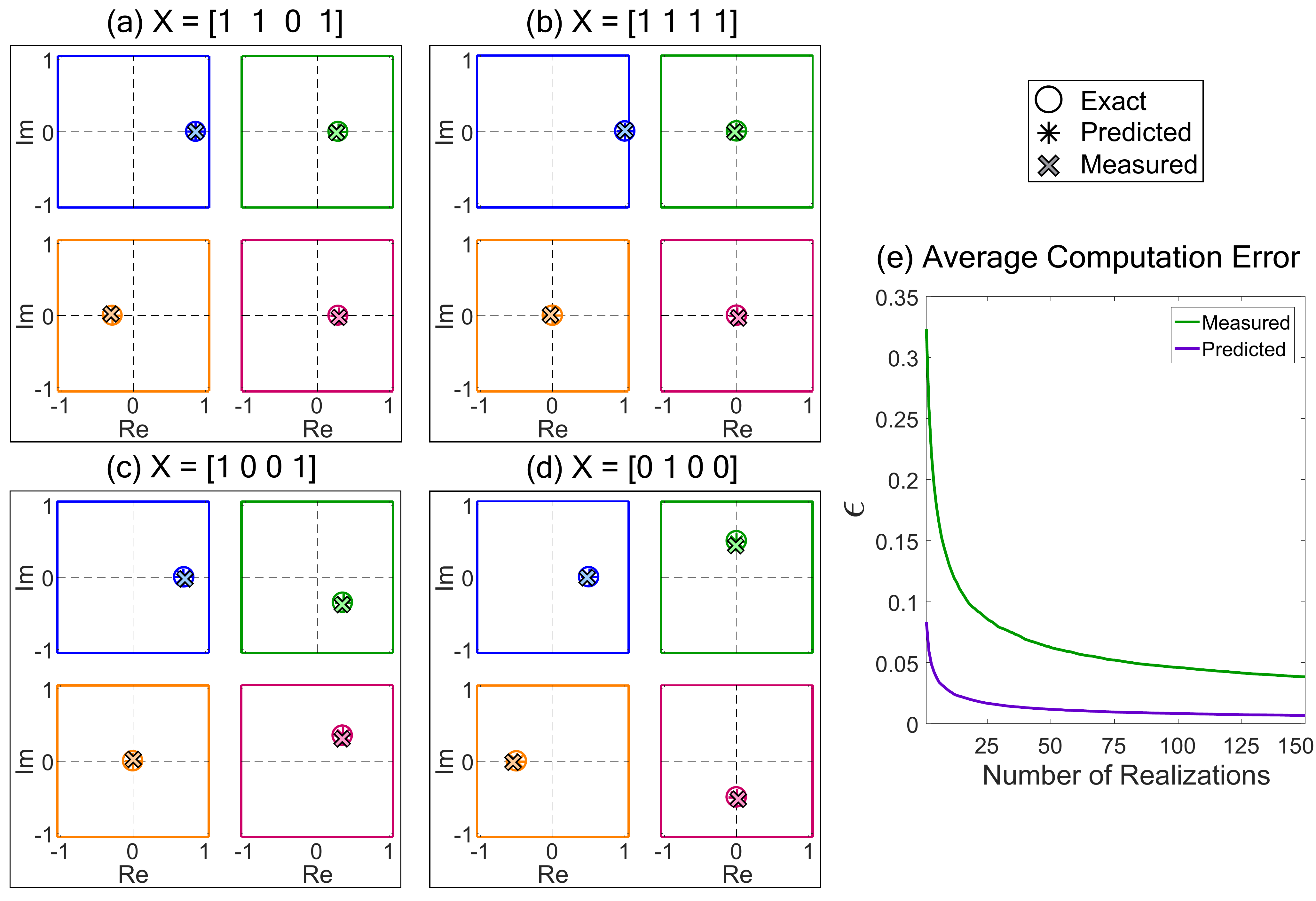}
	\caption{Experimentally obtained computation results for the complex-valued $4\times 4$ discrete Fourier transform operation. 
    (a-d) Results for the different indicated input vectors $\mathrm{X}$, normalized such that $\Sigma_j \lvert Y_j\rvert ^2 = 1$ for easy comparison. The experimentally measured values, $\mathrm{Y}^{\mathrm{meas}}$ (cross symbol), are shown together with the theoretically expected results for $\mathrm{Y}=\mathbf{G}\mathrm{X}$, $\mathrm{Y}^{\mathrm{exact}}$ (circle symbol), as well as the results predicted by the measured Impact Matrix (IM), $\mathrm{Y}^{\mathrm{pred}}$ (star symbol); cf.~legend. 
    (e) Dependence of the average computation error, $\epsilon = \langle \lvert Y^{\mathrm{exact}}_j - Y^{\mathrm{meas}}_j \rvert\rangle_j$ (green line), averaged over all possible $\mathrm{X}$, on the number of realizations over which the computation outcome was averaged. We contrast this with the average predicted error, $\epsilon' = \langle \lvert Y^{\mathrm{exact}}_j - Y^{\mathrm{pred}}_j \rvert\rangle_j$ (purple line), where $\mathrm{Y}^{\mathrm{pred}}$ is the expected computation outcome assuming the IM is exact.}
	\label{fig_results}
	\end{center}
\end{figure}

In Fig.~\ref{fig_results}(a-d) we present our experimentally obtained computation results $\mathrm{Y}$ for the discrete Fourier transform operation on four different input vectors $\mathrm{X}$, based on an ensemble average of $\mathrm{Y}^{\mathrm{meas}}$ over $150$ realizations ($30$ mode-stirrer positions, $5$ independent working frequencies). We observe an excellent agreement with the theoretically expected results, the average computation error $\epsilon = \langle \lvert Y^{\mathrm{exact}}_j - Y^{\mathrm{meas}}_j \rvert\rangle_j$ being $3.8\%$ (averaged over all possible input vectors $\mathrm{X}$). 

The dependence of $\epsilon$ on the number of realizations displayed in Fig.~\ref{fig_results}(e) confirms that the inaccuracy due to (consciously) neglecting the long-range correlations in the IM can indeed be averaged out successfully by ensemble averaging over different realizations. A further illustration of this effect based on experimental data is provided in Fig.~S2 of Ref.~[63]. 

Incidentally, we note that ensemble averaging is also extremely useful in case of very limited wave front control. For instance, in our experiment each of the four wave front segments is made up of only $22$ phase-binary pixels. As seen by the purple curve in Fig.~4(e), this is on average not enough control to perfectly tame the contributions from all input segments to all observation points. 
For a single realization, an entry of the predicted computation outcome $Y_j^{\mathrm{pred}}$, which does not suffer from the long-range correlations since it assumes the IM is exact, is not exactly equal to the exact result $Y_j^{\mathrm{exact}}$ expected for the operation. 
Once again, we can interpret this effect of imperfect wave front taming as a random realization-dependent phasor $\delta'_j$: $Y_j^{\mathrm{pred}} =Y_j^{\mathrm{exact}} + \delta'_j $. By the same token as before, $\langle \delta'_j\rangle _{\mathrm{realizations}} = 0$; indeed, as shown by the purple line in Fig.~\ref{fig_results}(e), $\epsilon' = \langle |Y_j^{\mathrm{exact}} - Y_j^{\mathrm{pred}}|\rangle_j$ is reduced from $8.3\%$ to $0.7\%$ by ensemble averaging. Hence, the idea of ensemble averaging serves two purposes in our case.

The presented  results not only constitute a proof-of-principle \textit{experimental} implementation of our proposed scheme for WBAC with a random medium, but allow us moreover to conclude that it takes nothing but standard indoor wireless communication infrastructure and a simple tunable metasurface to perform accurate analog computation with reverberating microwaves.
While we could have used any complex-valued matrix for our demonstration, we deliberately selected the (discrete) Fourier transform operation. Besides its key role in all areas of physics and engineering, it also facilitates interesting interpretations of our experimental results. 
Complex media in conjunction with WFS in optics, or time reversal in acoustics, have repeatedly been described as ``opaque lens'' or ``scattering lens'' in the literature \cite{NatPhotReview,Mosk_vis_subwavelength_foc_byWS}, alluding to the most basic property of a traditional lens: it focuses light in geometrical optics. In our present work, we use a complex medium to Fourier transform an impinging wave front, such that it becomes possible to extend the analogy from the ray picture to the properties of a traditional lens in the realm of ``wave'' optics. 
While a lens in optics is a standard component, transposed to the microwave domain our concept unveils yet more of its intriguing properties: (i) lenses are not as widely-used; (ii) reflect-array and antennas, mimicking input and output of the Fourier transform, are not aligned in any way but placed randomly and the medium carrying out the operation is \textit{around} rather than between inputs and outputs. 
Essentially, by configuring the reflect-array according to the previously established library for a given input $\mathrm{X}$, we create a cavity geometry in which the transmission between the selected antenna pairs corresponds to the computation output $\mathrm{Y}$. 

\subsection{Scalability of Resolution and Operation Size}\label{scalability}

Typical concerns with WBAC schemes include questions about the achievable resolution and system size. 
%
%
%
In the realm of artifical neural networks (ANNs) that constitutes as aforementioned an important potential field of application for WBAC, limited resolution is in fact by no means an insurmountable barrier. The accuracy loss of 8-bit artificial neural networks in comparison with 32-bit computers was shown to be only $0.4\%$ on standard tasks \cite{EfficientDNN}; moreover, if the ANN has been specifically designed for low resolution, it can achieve similar results with as little as 1-bit or 2-bit resolution.

If we were to bluntly apply our previous procedure with the same physical system to a significantly larger computational operation, identifying appropriate configurations for each segment would become doubly more difficult: (i) given the fixed reflect-array size, each wave front segment would consist of even fewer pixels, further reducing the available degrees of freedom; (ii) each segment configuration would have to simultaneously satisfy even more constraints, since there are also more observation points due to the larger size of the output vector $\mathrm{Y}$. Averaging over more realizations could cushion these adverse effects to some extent, but there are more elegant solutions.

In the following, we extend our previous experiment with 1-bit input resolution and 4-entry input size to a system with more than 4-bit input resolution and a 16-entry input size --- using exactly the same physical system. To access a better resolution, we simply perform the numerical optimization step outlined in Section~\ref{ExperimentalProcedure} for a larger range of values of $\alpha$. Note that this is only necessary because of the phase-binary only control of our metasurface reflect-array. To access a larger system size, we exploit the extremely low time cost of WBAC. 
Indeed, since the computation in our cavity implementation is carried out as the stationary wave field is established, the only potentially limiting factors are the field's transient duration and the duration of the measurement itself that is required to read off the computation result. 
We simply break our computational task down into several smaller ones that fit our physical system's size. For ANNs, rather than processing an input signal such as an image instantaneously, it has already become common practice to feed the input into the system via multiple patches one after the other --- even without the benefit of ultrafast computation times as in WBAC \cite{caffe}. Here, we demonstrate a time-sequential WBAC version that exploits the linearity of the computation operation $\mathrm{Y}=\mathbf{G}\mathrm{X}$ in which the entries of $\mathrm{Y}$ are independent from each other. Thereby, we compute one entry of $\mathrm{Y}$ after the other, such that we convert one $ \mathcal{N}\times \mathcal{N}$ operation into $ \mathcal{N}$ $1\times \mathcal{N}$ operations (see Fig.~S3 in Ref.~[63] for an illustration) that are performed one after the other. This is possible thanks to the reconfigurability of our wave-based analog computation unit --- a fundamental advantage of our proposed system over metamaterial computation units fabricated for a single specific task. 

In principle, this idea enables the computation of a subset of output entries simultaneously, in the following one at a time, depending on the size of the desired operation and the size of the available physical system. Even if we compute ``only'' one output entry at a time, the random medium still plays a crucial role in mixing all the tamed contributions from the different input segments to that single output observation point; the numerical optimization of wave front segment configurations during the one-off initiation step of our WBAC scheme is still more demanding than a simple focusing experiment, since we still need to accurately mimic one entire row of $\mathbf{G}$. In other words, we still need to accurately tame as many random walks as there are input entries, rather than only aligning the steps of a single walk.

\begin{figure}[t]
	\begin{center}
\includegraphics [width=\columnwidth] {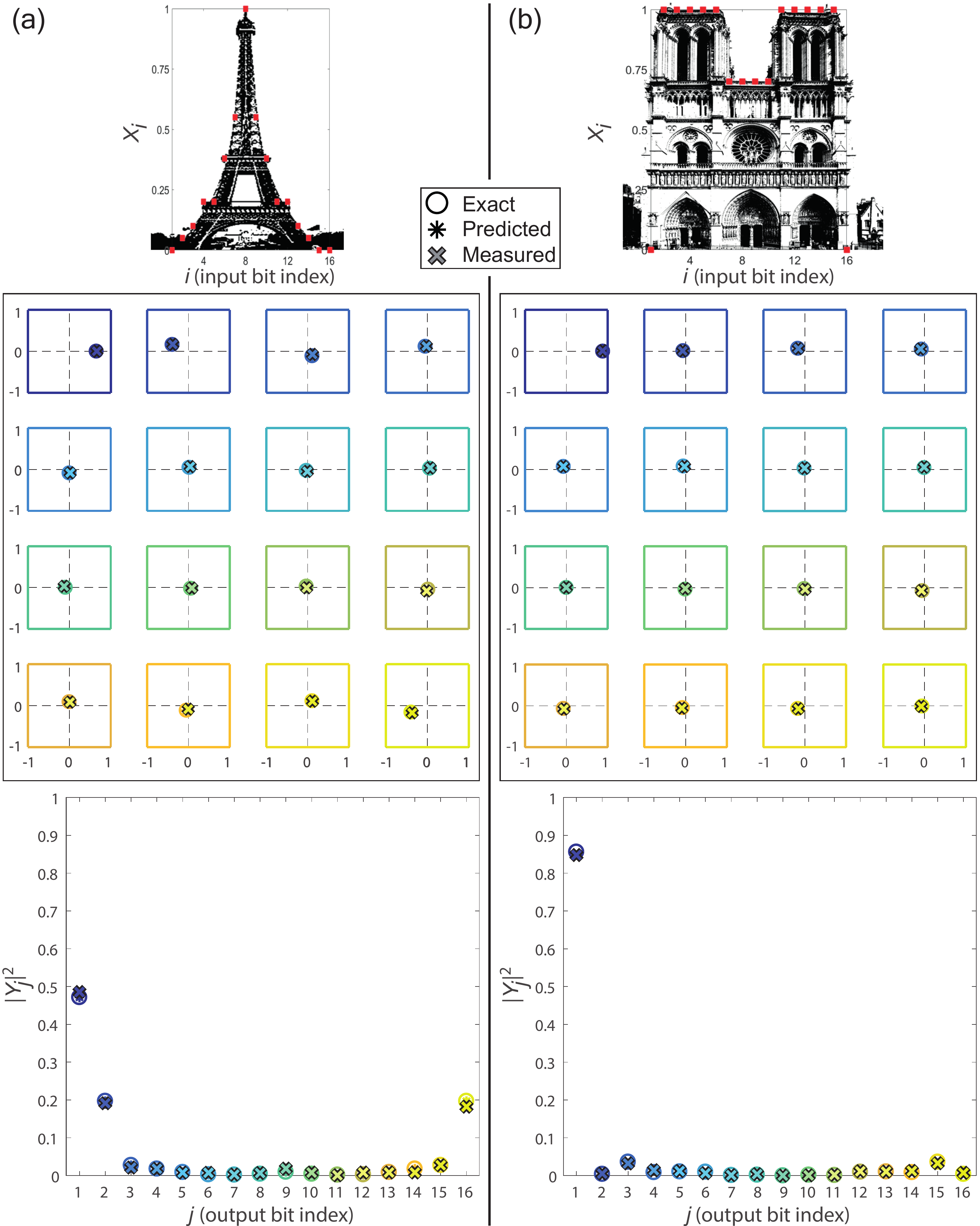}
	\caption{Demonstration of sequential wave-based analog computation. We recofigure the computation unit to compute one output at a time, which allows us to tackle larger computational tasks. The figure shows the computed Fourier transform of the silhouette of two Parisian monuments: (a) Eiffel Tower and (b) Notre Dame. We display the experimentally obtained results in the complex plane as well as in terms of the intensities. The exact results are also indicated, see legend.} 
	\label{fig_seq}
	\end{center}
\end{figure}

We present in Fig.~\ref{fig_seq} the experimentally obtained Fourier transform of the silhouette of two Parisian monuments. To faithfully reproduce the silhouettes, we used an input resolution of more than 4-bits, clearly superior to the 1-bit resolution from Fig.~\ref{fig_results}. We once again ensemble averaged over $150$ realizations of this $16\times 16$ complex-valued computation (see Section~D in Ref.~[63] for details on $\mathbf{G}$ and further examples). 
The obtained results are in good agreement with the exact ones, showing that the Eiffel Tower creates a richer spectrum than Notre Dame since it has more structural details.

Note that the presented multiplexing in time could in principle also be implemented in space with $ \mathcal{N}$ identical WBAC units. This might be of interest for optical hardware accelerators to circumvent a potential upper limit on computation power per system (limited degrees of freedom in complex medium, limited number of pixels on the WFS device) to maintain the ultrafast computation time.

WBAC also offers further exciting opportunities for parallelization of computational tasks. One can for instance imagine to operate our scheme at two or more clearly distinct working frequencies within the same cavity equipped with one metasuraces per working frequency, each metasurface operating within a narrow band around its corresponding working frequency. At a given working frequencies, the other metasurfaces not operating at that working frequencies would simply act like a metal wall without any modulating effect.

\section{Discussion and outlook}

\textit{Practical, stable and efficient cavity implementation.} --- 
The size of the discussed cavity implementation can conveniently be scaled down by operating at a higher frequency. For example, operating in the K-band around $20\ \mathrm{GHz}$ would reduce the volume of the previous experiment by a factor of $10^{-3}$, yielding cavity dimensions that can potentially even be integrated into data center racks. 
Mode-stirring can conveniently be performed in an all-electronic manner, by adding more metasurface pixels dedicated to mode-stirring; putting these pixels into different random configurations yields different realizations \cite{CEM2018,ramiroEMC}. This provides very fast access to more independent configurations than a mechanical mode-stirrer. The different realizations over which the computation outcomes are averaged can be predefined such that all initiation steps are performed before the actual computation phase.
The metasurface design can be upgraded, for instance, for linear phase control with a varactor-populated mushroom-structure as in Ref.~\cite{TimDMA}. 
The cavity constitutes a very stable enclosure that is completely shielded from the outside. The stability implies that once measured, the IM would not loose validity over time. The shielding allows for a fantastic signal-to-noise ratio. The very low noise level would almost eliminate any limits on resolution caused by effective truncation errors originating from noise. Moreover, the low noise level is very favorable for high-speed low-power operation, as discussed below.

\textit{Speed.} --- The fundamental advantage in operation speed over traditional processors originates from the fact that a WBAC unit performs a task in one go that requires many consecutive operations on a digital computer. For instance, an $\mathcal{N}\times\mathcal{N}$ matrix multiplication that we perform in a single run requires in general $O(\mathcal{N}^2)$ digital operations. Traditional computers, having clock rates on the order of one $\mathrm{GHz}$, can perform $\sim 10^9$ operations per second. In our cavity scheme, the limiting factor is the time it takes for the wave field to reach equilibrium, on the order of $\tau \sim 1/\Delta f_{\mathrm{corr}} \sim  5\ \mathrm{ns}$ for the K-band version, since measurements even at $\mathrm{GHz}$ rates are possible with standard technology --- aided by the low-noise nature of the shielded cavity. Accounting for the need for ensemble averaging, we think one entire computation can be performed on the order of $1 \ \mathrm{\mu s}$. The actual averaging over realizations could be performed directly on the detector, similarly to a camera that integrates the incident light over 
desired time intervals. 
Subtracting the constant cavity contribution $U_j$ from each observation point, given prior calibration, could probably be integrated into the measurement process, too, but in any case this is only of $O(\mathcal{N})$ complexity. Thus, we expect to be able to perform the equivalent of $\mathcal{N}^2\times 10^{6}$ digital operations per second. Clearly, the larger $ \mathcal{N}$ gets the higher is the advantage of the WBAC scheme. The break-even is expected for operation sizes on the order of $\mathcal{N} \sim 30$.

\textit{Energy efficiency.} --- A traditional computer uses at least $~100\ \mathrm{pJ}$ per operation \cite{MITonn,horowitz20141}. The power consumed by our cavity WBAC scheme is dominated by the feeding circuits of the metasurface pixels: $25\ \mathrm{\mu W}$ per pixel (see Ref.~\cite{SMM_PoC} and Section~E in Ref.~[63] for further details). This number could be reduced by taking power consumption as one of the design criteria of the metasurface. Using $\mathcal{N}^2 \times 10^{-6} \mathrm{\ W}$ as power needed to perform an $\mathcal{N}\times\mathcal{N}$ operation then yields an estimated energy consumption of $1\ \mathrm{pJ}$ per equivalent digital operation. Thus, the scheme has the potential to compete or even outperform a traditional computer in terms of energy efficiency.

\textit{Perspectives in optics.} --- An optical implementation of our scheme in a transmission geometry as in Fig.~\ref{fig0}(b) holds several promises, notably that the IM would not have any long-range correlations (and thus no intrinsic need for ensemble averaging), the operation speed in a transmission geometry would be limited only by the detection rate which can be on the order of $100\ \mathrm{GHz}$ with commercially available devices \cite{MITonn}, and the encoding of the input vector would be outside the WBAC system such that the latter would be completely passive during a given computation, not consuming any energy. The latter implies that the advantage in energy efficiency over an electronic computer would scale as $\mathcal{N}^2$.
Ultimately, these promises will have to be traded off against less favorable noise levels, resulting effective truncation errors, and the difficulty of accessing the computation outputs' phase information.

\section{Conclusion}\label{Conclusion}
We have demonstrated \textit{experimentally} that any random medium can be employed as reconfigurable wave-based analog computation unit, subject to appropriate wave front shaping. 
Our proposal removes the hitherto crux of implementing wave-based analog computation in practice: the need for intricate metamaterial designs that are difficult (if not impossible so far) to fabricate. Moreover, our unit is reconfigurable to perform a different operation, without any additional calibration measurements. Our experimental demonstration leveraging infrastructure analogous to standard wireless indoor communication systems illustrates the applicability of our principle to all types of wave phenomena and random media, as well as the ease of practical implementation.

Our scheme's practicality and reconfigurability may play important roles in the renaissance of specific-purpose computation, heralded in particular by the rapidly expanding utilization of artificial neural networks, demanding high-speed low-power processing of big data. Reconfigurability of artificial neural networks is key to enable transfer learning as well as to correct errors of previous training phases \cite{TransferLearningPreciseCitation}. 
Future versions of our concept can be extended to performing non-linear operations \cite{NL_Sievenpiper} and thereby become themselves a \textit{complete}, tailored hardware architecture for implementations of artificial neural networks \cite{kernels,MITonn}. This may include cascading several reconfigurable cavities, possibly connected via non-linear elements. There may also be possibilities to leverage the intrinsic IM inter-element long-range correlations that are so far considered an obstacle.
Furthermore, an exciting avenue for our concept is an implementation using optical multimode waveguides offering a stable and energy-efficient random medium for which wave front shaping has become a well-mastered technique \cite{RotterPRX}.

\section*{Acknowledgments}
We thank S\'{e}bastien Popoff and Laurent Daudet for fruitful discussions. 
P.d.H. acknowledges funding from the French ``Minist\`{e}re de la D\'{e}fense, Direction G\'{e}n\'{e}rale de l'Armement''. 



\bigskip

The project was initiated, conceptualized, conducted and written up by P.d.H. Both authors thoroughly discussed the project.


\begin{thebibliography}{81}%
\makeatletter
\providecommand \@ifxundefined [1]{%
 \@ifx{#1\undefined}
}%
\providecommand \@ifnum [1]{%
 \ifnum #1\expandafter \@firstoftwo
 \else \expandafter \@secondoftwo
 \fi
}%
\providecommand \@ifx [1]{%
 \ifx #1\expandafter \@firstoftwo
 \else \expandafter \@secondoftwo
 \fi
}%
\providecommand \natexlab [1]{#1}%
\providecommand \enquote  [1]{``#1''}%
\providecommand \bibnamefont  [1]{#1}%
\providecommand \bibfnamefont [1]{#1}%
\providecommand \citenamefont [1]{#1}%
\providecommand \href@noop [0]{\@secondoftwo}%
\providecommand \href [0]{\begingroup \@sanitize@url \@href}%
\providecommand \@href[1]{\@@startlink{#1}\@@href}%
\providecommand \@@href[1]{\endgroup#1\@@endlink}%
\providecommand \@sanitize@url [0]{\catcode `\\12\catcode `\$12\catcode
  `\&12\catcode `\#12\catcode `\^12\catcode `\_12\catcode `\%12\relax}%
\providecommand \@@startlink[1]{}%
\providecommand \@@endlink[0]{}%
\providecommand \url  [0]{\begingroup\@sanitize@url \@url }%
\providecommand \@url [1]{\endgroup\@href {#1}{\urlprefix }}%
\providecommand \urlprefix  [0]{URL }%
\providecommand \Eprint [0]{\href }%
\providecommand \doibase [0]{http://dx.doi.org/}%
\providecommand \selectlanguage [0]{\@gobble}%
\providecommand \bibinfo  [0]{\@secondoftwo}%
\providecommand \bibfield  [0]{\@secondoftwo}%
\providecommand \translation [1]{[#1]}%
\providecommand \BibitemOpen [0]{}%
\providecommand \bibitemStop [0]{}%
\providecommand \bibitemNoStop [0]{.\EOS\space}%
\providecommand \EOS [0]{\spacefactor3000\relax}%
\providecommand \BibitemShut  [1]{\csname bibitem#1\endcsname}%
\let\auto@bib@innerbib\@empty
\bibitem [{\citenamefont {Solli}\ and\ \citenamefont
  {Jalali}(2015)}]{RevJalali}%
  \BibitemOpen
  \bibfield  {author} {\bibinfo {author} {\bibfnamefont {D.~R.}\ \bibnamefont
  {Solli}}\ and\ \bibinfo {author} {\bibfnamefont {B.}~\bibnamefont {Jalali}},\
  }\bibfield  {title} {\enquote {\bibinfo {title} {Analog optical computing},}\
  }\href@noop {} {\bibfield  {journal} {\bibinfo  {journal} {Nat. Photonics}\
  }\textbf {\bibinfo {volume} {9}},\ \bibinfo {pages} {704} (\bibinfo {year}
  {2015})}\BibitemShut {NoStop}%
\bibitem [{\citenamefont {Moore}(1965)}]{moore}%
  \BibitemOpen
  \bibfield  {author} {\bibinfo {author} {\bibfnamefont {G.~E.}\ \bibnamefont
  {Moore}},\ }\bibfield  {title} {\enquote {\bibinfo {title} {Cramming more
  components onto integrated circuits},}\ }\href@noop {} {\bibfield  {journal}
  {\bibinfo  {journal} {Electronics}\ }\textbf {\bibinfo {volume} {38}},\
  \bibinfo {pages} {114--117} (\bibinfo {year} {1965})}\BibitemShut {NoStop}%
\bibitem [{\citenamefont {Esmaeilzadeh}\ \emph {et~al.}(2011)\citenamefont
  {Esmaeilzadeh}, \citenamefont {Blem}, \citenamefont {Amant}, \citenamefont
  {Sankaralingam},\ and\ \citenamefont {Burger}}]{DarkSilicon}%
  \BibitemOpen
  \bibfield  {author} {\bibinfo {author} {\bibfnamefont {H.}~\bibnamefont
  {Esmaeilzadeh}}, \bibinfo {author} {\bibfnamefont {E.}~\bibnamefont {Blem}},
  \bibinfo {author} {\bibfnamefont {R.~St.}\ \bibnamefont {Amant}}, \bibinfo
  {author} {\bibfnamefont {K.}~\bibnamefont {Sankaralingam}}, \ and\ \bibinfo
  {author} {\bibfnamefont {D.}~\bibnamefont {Burger}},\ }\bibfield  {title}
  {\enquote {\bibinfo {title} {Dark silicon and the end of multicore
  scaling},}\ }in\ \href@noop {} {\emph {\bibinfo {booktitle} {38th Annual
  International Symposium on Computer Architecture (ISCA)}}}\ (\bibinfo
  {organization} {IEEE},\ \bibinfo {year} {2011})\ pp.\ \bibinfo {pages}
  {365--376}\BibitemShut {NoStop}%
\bibitem [{\citenamefont {Esser}\ \emph {et~al.}(2016)\citenamefont {Esser},
  \citenamefont {Merolla}, \citenamefont {Arthur}, \citenamefont {Cassidy},
  \citenamefont {Appuswamy}, \citenamefont {Andreopoulos}, \citenamefont
  {Berg}, \citenamefont {McKinstry}, \citenamefont {Melano}, \citenamefont
  {Barch}, \citenamefont {di~Nolfo}, \citenamefont {Datta}, \citenamefont
  {Amir}, \citenamefont {Taba}, \citenamefont {Flickner},\ and\ \citenamefont
  {Modha}}]{TrueNorth}%
  \BibitemOpen
  \bibfield  {author} {\bibinfo {author} {\bibfnamefont {S.~K.}\ \bibnamefont
  {Esser}}, \bibinfo {author} {\bibfnamefont {P.~A.}\ \bibnamefont {Merolla}},
  \bibinfo {author} {\bibfnamefont {J.~V.}\ \bibnamefont {Arthur}}, \bibinfo
  {author} {\bibfnamefont {A.~S.}\ \bibnamefont {Cassidy}}, \bibinfo {author}
  {\bibfnamefont {R.}~\bibnamefont {Appuswamy}}, \bibinfo {author}
  {\bibfnamefont {A.}~\bibnamefont {Andreopoulos}}, \bibinfo {author}
  {\bibfnamefont {D.~J.}\ \bibnamefont {Berg}}, \bibinfo {author}
  {\bibfnamefont {J.~L.}\ \bibnamefont {McKinstry}}, \bibinfo {author}
  {\bibfnamefont {T.}~\bibnamefont {Melano}}, \bibinfo {author} {\bibfnamefont
  {D.~R.}\ \bibnamefont {Barch}}, \bibinfo {author} {\bibfnamefont
  {C.}~\bibnamefont {di~Nolfo}}, \bibinfo {author} {\bibfnamefont
  {P.}~\bibnamefont {Datta}}, \bibinfo {author} {\bibfnamefont
  {A.}~\bibnamefont {Amir}}, \bibinfo {author} {\bibfnamefont {B.}~\bibnamefont
  {Taba}}, \bibinfo {author} {\bibfnamefont {M.~D.}\ \bibnamefont {Flickner}},
  \ and\ \bibinfo {author} {\bibfnamefont {D.~S.}\ \bibnamefont {Modha}},\
  }\bibfield  {title} {\enquote {\bibinfo {title} {Convolutional networks for
  fast, energy-efficient neuromorphic computing},}\ }\href@noop {} {\bibfield  {journal} {\bibinfo  {journal} {Proc.
  Natl. Acad. Sci. U.S.A.}\ }\textbf {\bibinfo {volume} {113}},\ \bibinfo
  {pages} {11441--11446} (\bibinfo {year} {2016})}\BibitemShut {NoStop}%
\bibitem [{\citenamefont {Graves}\ \emph {et~al.}(2016)\citenamefont {Graves},
  \citenamefont {Wayne}, \citenamefont {Reynolds}, \citenamefont {Harley},
  \citenamefont {Danihelka}, \citenamefont {Grabska-Barwi{\'n}ska},
  \citenamefont {Colmenarejo}, \citenamefont {Grefenstette}, \citenamefont
  {Ramalho}, \citenamefont {Agapiou} \emph {et~al.}}]{GoogleTPU}%
  \BibitemOpen
  \bibfield  {author} {\bibinfo {author} {\bibfnamefont {A.}~\bibnamefont
  {Graves}}, \bibinfo {author} {\bibfnamefont {G.}~\bibnamefont {Wayne}},
  \bibinfo {author} {\bibfnamefont {M.}~\bibnamefont {Reynolds}}, \bibinfo
  {author} {\bibfnamefont {T.}~\bibnamefont {Harley}}, \bibinfo {author}
  {\bibfnamefont {I.}~\bibnamefont {Danihelka}}, \bibinfo {author}
  {\bibfnamefont {A.}~\bibnamefont {Grabska-Barwi{\'n}ska}}, \bibinfo {author}
  {\bibfnamefont {S.~G.}\ \bibnamefont {Colmenarejo}}, \bibinfo {author}
  {\bibfnamefont {E.}~\bibnamefont {Grefenstette}}, \bibinfo {author}
  {\bibfnamefont {T.}~\bibnamefont {Ramalho}}, \bibinfo {author} {\bibfnamefont
  {J.}~\bibnamefont {Agapiou}},  \emph {et~al.},\ }\bibfield  {title} {\enquote
  {\bibinfo {title} {Hybrid computing using a neural network with dynamic
  external memory},}\ }\href@noop {} {\bibfield  {journal} {\bibinfo  {journal}
  {Nature}\ }\textbf {\bibinfo {volume} {538}},\ \bibinfo {pages} {471}
  (\bibinfo {year} {2016})}\BibitemShut {NoStop}%
\bibitem [{\citenamefont {Goodman}(1996)}]{GoodmanBookFourierOptics}%
  \BibitemOpen
  \bibfield  {author} {\bibinfo {author} {\bibfnamefont {J.}~\bibnamefont
  {Goodman}},\ }\href@noop {} {\emph {\bibinfo {title} {Introduction to Fourier
  Optics}}}\ (\bibinfo  {publisher} {McGraw-hill},\ \bibinfo {year}
  {1996})\BibitemShut {NoStop}%
\bibitem [{\citenamefont {Reck}\ \emph {et~al.}(1994)\citenamefont {Reck},
  \citenamefont {Zeilinger}, \citenamefont {Bernstein},\ and\ \citenamefont
  {Bertani}}]{reck1994experimental}%
  \BibitemOpen
  \bibfield  {author} {\bibinfo {author} {\bibfnamefont {M.}~\bibnamefont
  {Reck}}, \bibinfo {author} {\bibfnamefont {A.}~\bibnamefont {Zeilinger}},
  \bibinfo {author} {\bibfnamefont {H.~J.}\ \bibnamefont {Bernstein}}, \ and\
  \bibinfo {author} {\bibfnamefont {P.}~\bibnamefont {Bertani}},\ }\bibfield
  {title} {\enquote {\bibinfo {title} {Experimental realization of any discrete
  unitary operator},}\ }\href@noop {} {\bibfield  {journal} {\bibinfo
  {journal} {Phys. Rev. Lett.}\ }\textbf {\bibinfo {volume} {73}},\ \bibinfo
  {pages} {58} (\bibinfo {year} {1994})}\BibitemShut {NoStop}%
\bibitem [{\citenamefont {Zhu}\ \emph {et~al.}(2017)\citenamefont {Zhu},
  \citenamefont {Zhou}, \citenamefont {Lou}, \citenamefont {Ye}, \citenamefont
  {Qiu}, \citenamefont {Ruan},\ and\ \citenamefont {Fan}}]{FanNatComm}%
  \BibitemOpen
  \bibfield  {author} {\bibinfo {author} {\bibfnamefont {T.}~\bibnamefont
  {Zhu}}, \bibinfo {author} {\bibfnamefont {Y.}~\bibnamefont {Zhou}}, \bibinfo
  {author} {\bibfnamefont {Y.}~\bibnamefont {Lou}}, \bibinfo {author}
  {\bibfnamefont {H.}~\bibnamefont {Ye}}, \bibinfo {author} {\bibfnamefont
  {M.}~\bibnamefont {Qiu}}, \bibinfo {author} {\bibfnamefont {Z.}~\bibnamefont
  {Ruan}}, \ and\ \bibinfo {author} {\bibfnamefont {S.}~\bibnamefont {Fan}},\
  }\bibfield  {title} {\enquote {\bibinfo {title} {Plasmonic computing of
  spatial differentiation},}\ }\href@noop {} {\bibfield  {journal} {\bibinfo
  {journal} {Nat. Commun.}\ }\textbf {\bibinfo {volume} {8}},\ \bibinfo {pages}
  {15391} (\bibinfo {year} {2017})}\BibitemShut {NoStop}%
\bibitem [{\citenamefont {Silva}\ \emph {et~al.}(2014)\citenamefont {Silva},
  \citenamefont {Monticone}, \citenamefont {Castaldi}, \citenamefont {Galdi},
  \citenamefont {Al{\`u}},\ and\ \citenamefont {Engheta}}]{EnghetaScience}%
  \BibitemOpen
  \bibfield  {author} {\bibinfo {author} {\bibfnamefont {A.}~\bibnamefont
  {Silva}}, \bibinfo {author} {\bibfnamefont {F.}~\bibnamefont {Monticone}},
  \bibinfo {author} {\bibfnamefont {G.}~\bibnamefont {Castaldi}}, \bibinfo
  {author} {\bibfnamefont {V.}~\bibnamefont {Galdi}}, \bibinfo {author}
  {\bibfnamefont {A.}~\bibnamefont {Al{\`u}}}, \ and\ \bibinfo {author}
  {\bibfnamefont {N.}~\bibnamefont {Engheta}},\ }\bibfield  {title} {\enquote
  {\bibinfo {title} {Performing mathematical operations with metamaterials},}\
  }\href@noop {} {\bibfield  {journal} {\bibinfo  {journal} {Science}\ }\textbf
  {\bibinfo {volume} {343}},\ \bibinfo {pages} {160--163} (\bibinfo {year}
  {2014})}\BibitemShut {NoStop}%
\bibitem [{\citenamefont {Monticone}\ \emph {et~al.}(2013)\citenamefont
  {Monticone}, \citenamefont {Estakhri},\ and\ \citenamefont
  {Al{\`u}}}]{monticonePRL}%
  \BibitemOpen
  \bibfield  {author} {\bibinfo {author} {\bibfnamefont {F.}~\bibnamefont
  {Monticone}}, \bibinfo {author} {\bibfnamefont {N.~M.}\ \bibnamefont
  {Estakhri}}, \ and\ \bibinfo {author} {\bibfnamefont {A.}~\bibnamefont
  {Al{\`u}}},\ }\bibfield  {title} {\enquote {\bibinfo {title} {Full control of
  nanoscale optical transmission with a composite metascreen},}\ }\href@noop {}
  {\bibfield  {journal} {\bibinfo  {journal} {Phys. Rev. Lett.}\ }\textbf
  {\bibinfo {volume} {110}},\ \bibinfo {pages} {203903} (\bibinfo {year}
  {2013})}\BibitemShut {NoStop}%
\bibitem [{\citenamefont {Sihvola}(2014)}]{sihvolaScience}%
  \BibitemOpen
  \bibfield  {author} {\bibinfo {author} {\bibfnamefont {A.}~\bibnamefont
  {Sihvola}},\ }\bibfield  {title} {\enquote {\bibinfo {title} {Enabling
  optical analog computing with metamaterials},}\ }\href@noop {} {\bibfield
  {journal} {\bibinfo  {journal} {Science}\ }\textbf {\bibinfo {volume}
  {343}},\ \bibinfo {pages} {144--145} (\bibinfo {year} {2014})}\BibitemShut
  {NoStop}%
\bibitem [{\citenamefont {Miller}(2013)}]{miller2013self}%
  \BibitemOpen
  \bibfield  {author} {\bibinfo {author} {\bibfnamefont {D.~A.~B.}\
  \bibnamefont {Miller}},\ }\bibfield  {title} {\enquote {\bibinfo {title}
  {Self-configuring universal linear optical component},}\ }\href@noop {}
  {\bibfield  {journal} {\bibinfo  {journal} {Photon. Res.}\ }\textbf {\bibinfo
  {volume} {1}},\ \bibinfo {pages} {1--15} (\bibinfo {year}
  {2013})}\BibitemShut {NoStop}%
\bibitem [{\citenamefont {Carolan}\ \emph {et~al.}(2015)\citenamefont
  {Carolan}, \citenamefont {Harrold}, \citenamefont {Sparrow}, \citenamefont
  {Mart{\'\i}n-L{\'o}pez}, \citenamefont {Russell}, \citenamefont
  {Silverstone}, \citenamefont {Shadbolt}, \citenamefont {Matsuda},
  \citenamefont {Oguma}, \citenamefont {Itoh} \emph {et~al.}}]{ScienceInterf}%
  \BibitemOpen
  \bibfield  {author} {\bibinfo {author} {\bibfnamefont {J.}~\bibnamefont
  {Carolan}}, \bibinfo {author} {\bibfnamefont {C.}~\bibnamefont {Harrold}},
  \bibinfo {author} {\bibfnamefont {C.}~\bibnamefont {Sparrow}}, \bibinfo
  {author} {\bibfnamefont {E.}~\bibnamefont {Mart{\'\i}n-L{\'o}pez}}, \bibinfo
  {author} {\bibfnamefont {N.~J.}\ \bibnamefont {Russell}}, \bibinfo {author}
  {\bibfnamefont {J.~W.}\ \bibnamefont {Silverstone}}, \bibinfo {author}
  {\bibfnamefont {P.~J.}\ \bibnamefont {Shadbolt}}, \bibinfo {author}
  {\bibfnamefont {N.}~\bibnamefont {Matsuda}}, \bibinfo {author} {\bibfnamefont
  {M.}~\bibnamefont {Oguma}}, \bibinfo {author} {\bibfnamefont
  {M.}~\bibnamefont {Itoh}},  \emph {et~al.},\ }\bibfield  {title} {\enquote
  {\bibinfo {title} {Universal linear optics},}\ }\href@noop {} {\bibfield
  {journal} {\bibinfo  {journal} {Science}\ }\textbf {\bibinfo {volume}
  {349}},\ \bibinfo {pages} {711--716} (\bibinfo {year} {2015})}\BibitemShut
  {NoStop}%
\bibitem [{\citenamefont {Ribeiro}\ \emph {et~al.}(2016)\citenamefont
  {Ribeiro}, \citenamefont {Ruocco}, \citenamefont {Vanacker},\ and\
  \citenamefont {Bogaerts}}]{ribeiro2016demonstration}%
  \BibitemOpen
  \bibfield  {author} {\bibinfo {author} {\bibfnamefont {A.}~\bibnamefont
  {Ribeiro}}, \bibinfo {author} {\bibfnamefont {A.}~\bibnamefont {Ruocco}},
  \bibinfo {author} {\bibfnamefont {L.}~\bibnamefont {Vanacker}}, \ and\
  \bibinfo {author} {\bibfnamefont {W.}~\bibnamefont {Bogaerts}},\ }\bibfield
  {title} {\enquote {\bibinfo {title} {Demonstration of a 4$\times$4-port
  universal linear circuit},}\ }\href@noop {} {\bibfield  {journal} {\bibinfo
  {journal} {Optica}\ }\textbf {\bibinfo {volume} {3}},\ \bibinfo {pages}
  {1348--1357} (\bibinfo {year} {2016})}\BibitemShut {NoStop}%
\bibitem [{\citenamefont {Shen}\ \emph {et~al.}(2017)\citenamefont {Shen},
  \citenamefont {Harris}, \citenamefont {Skirlo}, \citenamefont {Prabhu},
  \citenamefont {Baehr-Jones}, \citenamefont {Hochberg}, \citenamefont {Sun},
  \citenamefont {Zhao}, \citenamefont {Larochelle}, \citenamefont {Englund},\
  and\ \citenamefont {Solja\v{c}i\'{c}}}]{MITonn}%
  \BibitemOpen
  \bibfield  {author} {\bibinfo {author} {\bibfnamefont {Y.}~\bibnamefont
  {Shen}}, \bibinfo {author} {\bibfnamefont {N.~C.}\ \bibnamefont {Harris}},
  \bibinfo {author} {\bibfnamefont {S.}~\bibnamefont {Skirlo}}, \bibinfo
  {author} {\bibfnamefont {M.}~\bibnamefont {Prabhu}}, \bibinfo {author}
  {\bibfnamefont {T.}~\bibnamefont {Baehr-Jones}}, \bibinfo {author}
  {\bibfnamefont {M.}~\bibnamefont {Hochberg}}, \bibinfo {author}
  {\bibfnamefont {X.}~\bibnamefont {Sun}}, \bibinfo {author} {\bibfnamefont
  {S.}~\bibnamefont {Zhao}}, \bibinfo {author} {\bibfnamefont {H.}~\bibnamefont
  {Larochelle}}, \bibinfo {author} {\bibfnamefont {D.}~\bibnamefont {Englund}},
  \ and\ \bibinfo {author} {\bibfnamefont {M.}~\bibnamefont
  {Solja\v{c}i\'{c}}},\ }\bibfield  {title} {\enquote {\bibinfo {title} {Deep
  learning with coherent nanophotonic circuits},}\ }\href@noop {} {\bibfield
  {journal} {\bibinfo  {journal} {Nat. Photonics}\ }\textbf {\bibinfo {volume}
  {11}},\ \bibinfo {pages} {441} (\bibinfo {year} {2017})}\BibitemShut
  {NoStop}%
\bibitem [{\citenamefont {Miller}(2017)}]{MillerInterfRev}%
  \BibitemOpen
  \bibfield  {author} {\bibinfo {author} {\bibfnamefont {D.~A.~B.}\
  \bibnamefont {Miller}},\ }\bibfield  {title} {\enquote {\bibinfo {title}
  {Silicon photonics: Meshing optics with applications},}\ }\href@noop {}
  {\bibfield  {journal} {\bibinfo  {journal} {Nat. Photonics}\ }\textbf
  {\bibinfo {volume} {11}},\ \bibinfo {pages} {403} (\bibinfo {year}
  {2017})}\BibitemShut {NoStop}%
\bibitem [{\citenamefont {Goodfellow}\ \emph {et~al.}(2016)\citenamefont
  {Goodfellow}, \citenamefont {Bengio},\ and\ \citenamefont
  {Courville}}]{DeepLearning}%
  \BibitemOpen
  \bibfield  {author} {\bibinfo {author} {\bibfnamefont {I.}~\bibnamefont
  {Goodfellow}}, \bibinfo {author} {\bibfnamefont {Y.}~\bibnamefont {Bengio}},
  \ and\ \bibinfo {author} {\bibfnamefont {A.}~\bibnamefont {Courville}},\
  }\href@noop {} {\emph {\bibinfo {title} {Deep learning}}}\ (\bibinfo
  {publisher} {MIT Press},\ \bibinfo {year} {2016})\BibitemShut {NoStop}%
\bibitem [{\citenamefont {Li}\ \emph {et~al.}(2017)\citenamefont {Li},
  \citenamefont {Kim}, \citenamefont {Luo}, \citenamefont {Li}, \citenamefont
  {Long},\ and\ \citenamefont {Sievenpiper}}]{NL_Sievenpiper}%
  \BibitemOpen
  \bibfield  {author} {\bibinfo {author} {\bibfnamefont {A.}~\bibnamefont
  {Li}}, \bibinfo {author} {\bibfnamefont {S.}~\bibnamefont {Kim}}, \bibinfo
  {author} {\bibfnamefont {Y.}~\bibnamefont {Luo}}, \bibinfo {author}
  {\bibfnamefont {Y.}~\bibnamefont {Li}}, \bibinfo {author} {\bibfnamefont
  {J.}~\bibnamefont {Long}}, \ and\ \bibinfo {author} {\bibfnamefont {D.~F.}\
  \bibnamefont {Sievenpiper}},\ }\bibfield  {title} {\enquote {\bibinfo {title}
  {High-power transistor-based tunable and switchable metasurface absorber},}\
  }\href@noop {} {\bibfield  {journal} {\bibinfo  {journal} {IEEE Trans.
  Microw. Theory Techn.}\ }\textbf {\bibinfo {volume} {65}},\ \bibinfo {pages}
  {2810--2818} (\bibinfo {year} {2017})}\BibitemShut {NoStop}%
\bibitem [{\citenamefont {Freund}(1990)}]{freund}%
  \BibitemOpen
  \bibfield  {author} {\bibinfo {author} {\bibfnamefont {I.}~\bibnamefont
  {Freund}},\ }\bibfield  {title} {\enquote {\bibinfo {title} {Looking through
  walls and around corners},}\ }\href@noop {} {\bibfield  {journal} {\bibinfo
  {journal} {Physica A}\ }\textbf {\bibinfo {volume} {168}},\ \bibinfo {pages}
  {49--65} (\bibinfo {year} {1990})}\BibitemShut {NoStop}%
\bibitem [{\citenamefont {{Fink}}(1997)}]{TR_fink}%
  \BibitemOpen
  \bibfield  {author} {\bibinfo {author} {\bibfnamefont {M.}~\bibnamefont
  {{Fink}}},\ }\bibfield  {title} {\enquote {\bibinfo {title} {Time reversed
  acoustics},}\ }\href@noop {} {\bibfield  {journal}
  {\bibinfo  {journal} {Phys. Today}\ }\textbf {\bibinfo {volume} {50}},\
  \bibinfo {pages} {34--40} (\bibinfo {year} {1997})}\BibitemShut {NoStop}%
\bibitem [{\citenamefont {Vellekoop}\ and\ \citenamefont
  {Mosk}(2007)}]{mosk_SLM}%
  \BibitemOpen
  \bibfield  {author} {\bibinfo {author} {\bibfnamefont {I.~M.}\ \bibnamefont
  {Vellekoop}}\ and\ \bibinfo {author} {\bibfnamefont {A.~P.}\ \bibnamefont
  {Mosk}},\ }\bibfield  {title} {\enquote {\bibinfo {title} {Focusing coherent
  light through opaque strongly scattering media},}\ }\href@noop {} {\bibfield
  {journal} {\bibinfo  {journal} {Opt. Lett.}\ }\textbf {\bibinfo {volume}
  {32}},\ \bibinfo {pages} {2309--2311} (\bibinfo {year} {2007})}\BibitemShut
  {NoStop}%
\bibitem [{\citenamefont {Mosk}\ \emph {et~al.}(2012)\citenamefont {Mosk},
  \citenamefont {Lagendijk}, \citenamefont {Lerosey},\ and\ \citenamefont
  {Fink}}]{NatPhotReview}%
  \BibitemOpen
  \bibfield  {author} {\bibinfo {author} {\bibfnamefont {A.~P.}\ \bibnamefont
  {Mosk}}, \bibinfo {author} {\bibfnamefont {A.}~\bibnamefont {Lagendijk}},
  \bibinfo {author} {\bibfnamefont {G.}~\bibnamefont {Lerosey}}, \ and\
  \bibinfo {author} {\bibfnamefont {M.}~\bibnamefont {Fink}},\ }\bibfield
  {title} {\enquote {\bibinfo {title} {Controlling waves in space and time for
  imaging and focusing in complex media},}\ }\href@noop {} {\bibfield
  {journal} {\bibinfo  {journal} {Nat. Photonics}\ }\textbf {\bibinfo {volume}
  {6}},\ \bibinfo {pages} {283--292} (\bibinfo {year} {2012})}\BibitemShut
  {NoStop}%
\bibitem [{\citenamefont {Rotter}\ and\ \citenamefont
  {Gigan}(2017)}]{RotterGigan}%
  \BibitemOpen
  \bibfield  {author} {\bibinfo {author} {\bibfnamefont {S.}~\bibnamefont
  {Rotter}}\ and\ \bibinfo {author} {\bibfnamefont {S.}~\bibnamefont {Gigan}},\
  }\bibfield  {title} {\enquote {\bibinfo {title} {Light fields in complex
  media: Mesoscopic scattering meets wave control},}\ }\href@noop {} {\bibfield
   {journal} {\bibinfo  {journal} {Rev. Mod. Phys.}\ }\textbf {\bibinfo
  {volume} {89}},\ \bibinfo {pages} {015005} (\bibinfo {year}
  {2017})}\BibitemShut {NoStop}%
\bibitem [{\citenamefont {Popoff}\ \emph {et~al.}(2010)\citenamefont {Popoff},
  \citenamefont {Lerosey}, \citenamefont {Carminati}, \citenamefont {Fink},
  \citenamefont {Boccara},\ and\ \citenamefont {Gigan}}]{popoff_prl}%
  \BibitemOpen
  \bibfield  {author} {\bibinfo {author} {\bibfnamefont {S.~M.}\ \bibnamefont
  {Popoff}}, \bibinfo {author} {\bibfnamefont {G.}~\bibnamefont {Lerosey}},
  \bibinfo {author} {\bibfnamefont {R.}~\bibnamefont {Carminati}}, \bibinfo
  {author} {\bibfnamefont {M.}~\bibnamefont {Fink}}, \bibinfo {author}
  {\bibfnamefont {A.~C.}\ \bibnamefont {Boccara}}, \ and\ \bibinfo {author}
  {\bibfnamefont {S.}~\bibnamefont {Gigan}},\ }\bibfield  {title} {\enquote
  {\bibinfo {title} {Measuring the transmission matrix in optics: an approach
  to the study and control of light propagation in disordered media},}\
  }\href@noop {} {\bibfield  {journal} {\bibinfo  {journal} {Phys. Rev. Lett.}\
  }\textbf {\bibinfo {volume} {104}},\ \bibinfo {pages} {100601} (\bibinfo
  {year} {2010})}\BibitemShut {NoStop}%
\bibitem [{\citenamefont {Aulbach}\ \emph {et~al.}(2011)\citenamefont
  {Aulbach}, \citenamefont {Gjonaj}, \citenamefont {Johnson}, \citenamefont
  {Mosk},\ and\ \citenamefont {Lagendijk}}]{aulbach_STF}%
  \BibitemOpen
  \bibfield  {author} {\bibinfo {author} {\bibfnamefont {J.}~\bibnamefont
  {Aulbach}}, \bibinfo {author} {\bibfnamefont {B.}~\bibnamefont {Gjonaj}},
  \bibinfo {author} {\bibfnamefont {P.~M.}\ \bibnamefont {Johnson}}, \bibinfo
  {author} {\bibfnamefont {A.~P.}\ \bibnamefont {Mosk}}, \ and\ \bibinfo
  {author} {\bibfnamefont {A.}~\bibnamefont {Lagendijk}},\ }\bibfield  {title}
  {\enquote {\bibinfo {title} {Control of light transmission through opaque
  scattering media in space and time},}\ }\href@noop {} {\bibfield  {journal} {\bibinfo  {journal}
  {Phys. Rev. Lett.}\ }\textbf {\bibinfo {volume} {106}},\ \bibinfo {pages}
  {103901} (\bibinfo {year} {2011})}\BibitemShut {NoStop}%
\bibitem [{\citenamefont {Katz}\ \emph {et~al.}(2011)\citenamefont {Katz},
  \citenamefont {Small}, \citenamefont {Bromberg},\ and\ \citenamefont
  {Silberberg}}]{katz_STF}%
  \BibitemOpen
  \bibfield  {author} {\bibinfo {author} {\bibfnamefont {O.}~\bibnamefont
  {Katz}}, \bibinfo {author} {\bibfnamefont {E.}~\bibnamefont {Small}},
  \bibinfo {author} {\bibfnamefont {Y.}~\bibnamefont {Bromberg}}, \ and\
  \bibinfo {author} {\bibfnamefont {Y.}~\bibnamefont {Silberberg}},\ }\bibfield
   {title} {\enquote {\bibinfo {title} {Focusing and compression of ultrashort
  pulses through scattering media},}\ }\href@noop {} {\bibfield  {journal}
  {\bibinfo  {journal} {Nat. Photonics}\ }\textbf {\bibinfo {volume} {5}},\
  \bibinfo {pages} {372--377} (\bibinfo {year} {2011})}\BibitemShut {NoStop}%
\bibitem [{\citenamefont {del Hougne}\ \emph
  {et~al.}(2016{\natexlab{a}})\citenamefont {del Hougne}, \citenamefont
  {Lemoult}, \citenamefont {Fink},\ and\ \citenamefont
  {Lerosey}}]{publikation3}%
  \BibitemOpen
  \bibfield  {author} {\bibinfo {author} {\bibfnamefont {P.}~\bibnamefont {del
  Hougne}}, \bibinfo {author} {\bibfnamefont {F.}~\bibnamefont {Lemoult}},
  \bibinfo {author} {\bibfnamefont {M.}~\bibnamefont {Fink}}, \ and\ \bibinfo
  {author} {\bibfnamefont {G.}~\bibnamefont {Lerosey}},\ }\bibfield  {title}
  {\enquote {\bibinfo {title} {Spatiotemporal wave front shaping in a microwave
  cavity},}\ }\href@noop {} {\bibfield  {journal} {\bibinfo  {journal} {Phys.
  Rev. Lett.}\ }\textbf {\bibinfo {volume} {117}},\ \bibinfo {pages} {134302}
  (\bibinfo {year} {2016}{\natexlab{a}})}\BibitemShut {NoStop}%
\bibitem [{\citenamefont {Vellekoop}\ \emph {et~al.}(2010)\citenamefont
  {Vellekoop}, \citenamefont {Lagendijk},\ and\ \citenamefont
  {Mosk}}]{mosk_disorder4perfectFOC}%
  \BibitemOpen
  \bibfield  {author} {\bibinfo {author} {\bibfnamefont {I.~M.}\ \bibnamefont
  {Vellekoop}}, \bibinfo {author} {\bibfnamefont {A.}~\bibnamefont
  {Lagendijk}}, \ and\ \bibinfo {author} {\bibfnamefont {A.~P.}\ \bibnamefont
  {Mosk}},\ }\bibfield  {title} {\enquote {\bibinfo {title} {Exploiting
  disorder for perfect focusing},}\ }\href@noop {} {\bibfield  {journal}
  {\bibinfo  {journal} {Nat. Photonics}\ }\textbf {\bibinfo {volume} {4}},\
  \bibinfo {pages} {320--322} (\bibinfo {year} {2010})}\BibitemShut {NoStop}%
\bibitem [{\citenamefont {Choi}\ \emph {et~al.}(2011)\citenamefont {Choi},
  \citenamefont {Yang}, \citenamefont {Fang-Yen}, \citenamefont {Kang},
  \citenamefont {Lee}, \citenamefont {Dasari}, \citenamefont {Feld},\ and\
  \citenamefont {Choi}}]{choi2011subwavelengthFOC}%
  \BibitemOpen
  \bibfield  {author} {\bibinfo {author} {\bibfnamefont {Y.}~\bibnamefont
  {Choi}}, \bibinfo {author} {\bibfnamefont {T.~D.}\ \bibnamefont {Yang}},
  \bibinfo {author} {\bibfnamefont {C.}~\bibnamefont {Fang-Yen}}, \bibinfo
  {author} {\bibfnamefont {P.}~\bibnamefont {Kang}}, \bibinfo {author}
  {\bibfnamefont {K.~J.}\ \bibnamefont {Lee}}, \bibinfo {author} {\bibfnamefont
  {R.~R.}\ \bibnamefont {Dasari}}, \bibinfo {author} {\bibfnamefont {M.~S.}\
  \bibnamefont {Feld}}, \ and\ \bibinfo {author} {\bibfnamefont
  {W.}~\bibnamefont {Choi}},\ }\bibfield  {title} {\enquote {\bibinfo {title}
  {Overcoming the diffraction limit using multiple light scattering in a highly
  disordered medium},}\ }\href@noop {} {\bibfield  {journal} {\bibinfo
  {journal} {Phys. Rev. Lett.}\ }\textbf {\bibinfo {volume} {107}},\ \bibinfo
  {pages} {023902} (\bibinfo {year} {2011})}\BibitemShut {NoStop}%
\bibitem [{\citenamefont {Park}\ \emph {et~al.}(2013)\citenamefont {Park},
  \citenamefont {Park}, \citenamefont {Yu}, \citenamefont {Park}, \citenamefont
  {Han}, \citenamefont {Shin}, \citenamefont {Ko}, \citenamefont {Nam},
  \citenamefont {Cho},\ and\ \citenamefont {Park}}]{park2013subwavelength}%
  \BibitemOpen
  \bibfield  {author} {\bibinfo {author} {\bibfnamefont {J.-H.}\ \bibnamefont
  {Park}}, \bibinfo {author} {\bibfnamefont {C.}~\bibnamefont {Park}}, \bibinfo
  {author} {\bibfnamefont {H.~S.}\ \bibnamefont {Yu}}, \bibinfo {author}
  {\bibfnamefont {J.}~\bibnamefont {Park}}, \bibinfo {author} {\bibfnamefont
  {S.}~\bibnamefont {Han}}, \bibinfo {author} {\bibfnamefont {J.}~\bibnamefont
  {Shin}}, \bibinfo {author} {\bibfnamefont {S.~H.}\ \bibnamefont {Ko}},
  \bibinfo {author} {\bibfnamefont {K.~T.}\ \bibnamefont {Nam}}, \bibinfo
  {author} {\bibfnamefont {Y.-H.}\ \bibnamefont {Cho}}, \ and\ \bibinfo
  {author} {\bibfnamefont {Y.~K.}\ \bibnamefont {Park}},\ }\bibfield  {title}
  {\enquote {\bibinfo {title} {Subwavelength light focusing using random
  nanoparticles},}\ }\href@noop {} {\bibfield  {journal} {\bibinfo  {journal}
  {Nat. Photonics}\ }\textbf {\bibinfo {volume} {7}},\ \bibinfo {pages}
  {454--458} (\bibinfo {year} {2013})}\BibitemShut {NoStop}%
\bibitem [{\citenamefont {Huisman}\ \emph {et~al.}(2015)\citenamefont
  {Huisman}, \citenamefont {Huisman}, \citenamefont {Wolterink}, \citenamefont
  {Mosk},\ and\ \citenamefont {Pinkse}}]{pinske}%
  \BibitemOpen
  \bibfield  {author} {\bibinfo {author} {\bibfnamefont {S.~R.}\ \bibnamefont
  {Huisman}}, \bibinfo {author} {\bibfnamefont {T.~J.}\ \bibnamefont
  {Huisman}}, \bibinfo {author} {\bibfnamefont {T.~A.~W.}\ \bibnamefont
  {Wolterink}}, \bibinfo {author} {\bibfnamefont {A.~P.}\ \bibnamefont {Mosk}},
  \ and\ \bibinfo {author} {\bibfnamefont {P.~W.~H.}\ \bibnamefont {Pinkse}},\
  }\bibfield  {title} {\enquote {\bibinfo {title} {Programmable multiport
  optical circuits in opaque scattering materials},}\ }\href@noop {} {\bibfield
   {journal} {\bibinfo  {journal} {Opt. Express}\ }\textbf {\bibinfo {volume}
  {23}},\ \bibinfo {pages} {3102--3116} (\bibinfo {year} {2015})}\BibitemShut
  {NoStop}%
\bibitem [{\citenamefont {Defienne}\ \emph {et~al.}(2016)\citenamefont
  {Defienne}, \citenamefont {Barbieri}, \citenamefont {Walmsley}, \citenamefont
  {Smith},\ and\ \citenamefont {Gigan}}]{hugo}%
  \BibitemOpen
  \bibfield  {author} {\bibinfo {author} {\bibfnamefont {H.}~\bibnamefont
  {Defienne}}, \bibinfo {author} {\bibfnamefont {M.}~\bibnamefont {Barbieri}},
  \bibinfo {author} {\bibfnamefont {I.~A.}\ \bibnamefont {Walmsley}}, \bibinfo
  {author} {\bibfnamefont {B.~J.}\ \bibnamefont {Smith}}, \ and\ \bibinfo
  {author} {\bibfnamefont {S.}~\bibnamefont {Gigan}},\ }\bibfield  {title}
  {\enquote {\bibinfo {title} {Two-photon quantum walk in a multimode fiber},}\
  }\href@noop {} {\bibfield  {journal} {\bibinfo  {journal} {Sci. Adv.}\
  }\textbf {\bibinfo {volume} {2}},\ \bibinfo {pages} {e1501054} (\bibinfo
  {year} {2016})}\BibitemShut {NoStop}%
\bibitem [{\citenamefont {Fickler}\ \emph {et~al.}(2017)\citenamefont
  {Fickler}, \citenamefont {Ginoya},\ and\ \citenamefont {Boyd}}]{fickler}%
  \BibitemOpen
  \bibfield  {author} {\bibinfo {author} {\bibfnamefont {R.}~\bibnamefont
  {Fickler}}, \bibinfo {author} {\bibfnamefont {M.}~\bibnamefont {Ginoya}}, \
  and\ \bibinfo {author} {\bibfnamefont {R.~W.}\ \bibnamefont {Boyd}},\
  }\bibfield  {title} {\enquote {\bibinfo {title} {Custom-tailored spatial mode
  sorting by controlled random scattering},}\ }\href@noop {} {\bibfield
  {journal} {\bibinfo  {journal} {Phys. Rev. B}\ }\textbf {\bibinfo {volume}
  {95}},\ \bibinfo {pages} {161108} (\bibinfo {year} {2017})}\BibitemShut
  {NoStop}%
\bibitem [{\citenamefont {Saade}\ \emph {et~al.}(2016)\citenamefont {Saade},
  \citenamefont {Caltagirone}, \citenamefont {Carron}, \citenamefont {Daudet},
  \citenamefont {Dr{\'e}meau}, \citenamefont {Gigan},\ and\ \citenamefont
  {Krzakala}}]{kernels}%
  \BibitemOpen
  \bibfield  {author} {\bibinfo {author} {\bibfnamefont {A.}~\bibnamefont
  {Saade}}, \bibinfo {author} {\bibfnamefont {F.}~\bibnamefont {Caltagirone}},
  \bibinfo {author} {\bibfnamefont {I.}~\bibnamefont {Carron}}, \bibinfo
  {author} {\bibfnamefont {L.}~\bibnamefont {Daudet}}, \bibinfo {author}
  {\bibfnamefont {A.}~\bibnamefont {Dr{\'e}meau}}, \bibinfo {author}
  {\bibfnamefont {S.}~\bibnamefont {Gigan}}, \ and\ \bibinfo {author}
  {\bibfnamefont {F.}~\bibnamefont {Krzakala}},\ }\bibfield  {title} {\enquote
  {\bibinfo {title} {Random projections through multiple optical scattering:
  Approximating kernels at the speed of light},}\ }in\ \href@noop {} {\emph
  {\bibinfo {booktitle} {IEEE International Conference on Acoustics, Speech and
  Signal Processing (ICASSP)}}}\ (\bibinfo {organization} {IEEE},\ \bibinfo
  {year} {2016})\ pp.\ \bibinfo {pages} {6215--6219}\BibitemShut {NoStop}%
\bibitem [{\citenamefont {Simon}\ \emph {et~al.}(2001)\citenamefont {Simon},
  \citenamefont {Moustakas}, \citenamefont {Stoytchev},\ and\ \citenamefont
  {Safar}}]{MIMO_PhysToday}%
  \BibitemOpen
  \bibfield  {author} {\bibinfo {author} {\bibfnamefont {S.~H.}\ \bibnamefont
  {Simon}}, \bibinfo {author} {\bibfnamefont {A.~L.}\ \bibnamefont
  {Moustakas}}, \bibinfo {author} {\bibfnamefont {M.}~\bibnamefont
  {Stoytchev}}, \ and\ \bibinfo {author} {\bibfnamefont {H.}~\bibnamefont
  {Safar}},\ }\bibfield  {title} {\enquote {\bibinfo {title} {Communication in
  a disordered world},}\ }\href@noop {} {\bibfield  {journal} {\bibinfo
  {journal} {Phys. Today}\ }\textbf {\bibinfo {volume} {54}},\ \bibinfo {pages}
  {38--43} (\bibinfo {year} {2001})}\BibitemShut {NoStop}%
\bibitem [{\citenamefont {Kim}\ \emph {et~al.}(2015)\citenamefont {Kim},
  \citenamefont {Choi}, \citenamefont {Choi}, \citenamefont {Yoon},\ and\
  \citenamefont {Choi}}]{choi2015WSforBioMed}%
  \BibitemOpen
  \bibfield  {author} {\bibinfo {author} {\bibfnamefont {M.}~\bibnamefont
  {Kim}}, \bibinfo {author} {\bibfnamefont {W.}~\bibnamefont {Choi}}, \bibinfo
  {author} {\bibfnamefont {Y.}~\bibnamefont {Choi}}, \bibinfo {author}
  {\bibfnamefont {C.}~\bibnamefont {Yoon}}, \ and\ \bibinfo {author}
  {\bibfnamefont {W.}~\bibnamefont {Choi}},\ }\bibfield  {title} {\enquote
  {\bibinfo {title} {Transmission matrix of a scattering medium and its
  applications in biophotonics},}\ }\href@noop {} {\bibfield  {journal}
  {\bibinfo  {journal} {Opt. Express}\ }\textbf {\bibinfo {volume} {23}},\
  \bibinfo {pages} {12648--12668} (\bibinfo {year} {2015})}\BibitemShut
  {NoStop}%
\bibitem [{\citenamefont {St{\"o}ckmann}(2007)}]{BookStockmann}%
  \BibitemOpen
  \bibfield  {author} {\bibinfo {author} {\bibfnamefont {H.-J.}\ \bibnamefont
  {St{\"o}ckmann}},\ }\href@noop {} {\emph {\bibinfo {title} {Quantum Chaos: An
  Introduction}}}\ (\bibinfo  {publisher} {Cambridge University Press},\
  \bibinfo {year} {2007})\BibitemShut {NoStop}%
\bibitem [{\citenamefont {Gollub}\ \emph {et~al.}(2017)\citenamefont {Gollub},
  \citenamefont {Yurduseven}, \citenamefont {Trofatter}, \citenamefont
  {Arnitz}, \citenamefont {Imani}, \citenamefont {Sleasman}, \citenamefont
  {Boyarsky}, \citenamefont {Rose}, \citenamefont {Pedross-Engel},
  \citenamefont {Odabasi} \emph {et~al.}}]{JonahSciRep}%
  \BibitemOpen
  \bibfield  {author} {\bibinfo {author} {\bibfnamefont {J.~N.}\ \bibnamefont
  {Gollub}}, \bibinfo {author} {\bibfnamefont {O.}~\bibnamefont {Yurduseven}},
  \bibinfo {author} {\bibfnamefont {K.~P.}\ \bibnamefont {Trofatter}}, \bibinfo
  {author} {\bibfnamefont {D.}~\bibnamefont {Arnitz}}, \bibinfo {author}
  {\bibfnamefont {M.~F.}\ \bibnamefont {Imani}}, \bibinfo {author}
  {\bibfnamefont {T.}~\bibnamefont {Sleasman}}, \bibinfo {author}
  {\bibfnamefont {M.}~\bibnamefont {Boyarsky}}, \bibinfo {author}
  {\bibfnamefont {A.}~\bibnamefont {Rose}}, \bibinfo {author} {\bibfnamefont
  {A.}~\bibnamefont {Pedross-Engel}}, \bibinfo {author} {\bibfnamefont
  {H.}~\bibnamefont {Odabasi}},  \emph {et~al.},\ }\bibfield  {title} {\enquote
  {\bibinfo {title} {Large metasurface aperture for millimeter wave
  computational imaging at the human-scale},}\ }\href@noop {} {\bibfield
  {journal} {\bibinfo  {journal} {Sci. Rep.}\ }\textbf {\bibinfo {volume}
  {7}},\ \bibinfo {pages} {42650} (\bibinfo {year} {2017})}\BibitemShut
  {NoStop}%
\bibitem [{\citenamefont {Sleasman}\ \emph {et~al.}(2016)\citenamefont
  {Sleasman}, \citenamefont {Imani}, \citenamefont {Gollub},\ and\
  \citenamefont {Smith}}]{TimDMA}%
  \BibitemOpen
  \bibfield  {author} {\bibinfo {author} {\bibfnamefont {T.}~\bibnamefont
  {Sleasman}}, \bibinfo {author} {\bibfnamefont {M.~F.}\ \bibnamefont {Imani}},
  \bibinfo {author} {\bibfnamefont {J.~N.}\ \bibnamefont {Gollub}}, \ and\
  \bibinfo {author} {\bibfnamefont {D.~R.}\ \bibnamefont {Smith}},\ }\bibfield
  {title} {\enquote {\bibinfo {title} {Microwave imaging using a disordered
  cavity with a dynamically tunable impedance surface},}\ }\href@noop {}
  {\bibfield  {journal} {\bibinfo  {journal} {Phys. Rev. Applied}\ }\textbf
  {\bibinfo {volume} {6}},\ \bibinfo {pages} {054019} (\bibinfo {year}
  {2016})}\BibitemShut {NoStop}%
\bibitem [{\citenamefont {Asefi}\ and\ \citenamefont
  {LoVetri}(2017)}]{winnipeg1}%
  \BibitemOpen
  \bibfield  {author} {\bibinfo {author} {\bibfnamefont {M.}~\bibnamefont
  {Asefi}}\ and\ \bibinfo {author} {\bibfnamefont {J.}~\bibnamefont
  {LoVetri}},\ }\bibfield  {title} {\enquote {\bibinfo {title} {Use of
  field-perturbing elements to increase nonredundant data for microwave imaging
  systems},}\ }\href@noop {} {\bibfield  {journal} {\bibinfo  {journal} {IEEE
  Trans. Microw. Theory Techn.}\ } (\bibinfo {year} {2017})}\BibitemShut
  {NoStop}%
\bibitem [{\citenamefont {del Hougne}\ \emph
  {et~al.}(2018{\natexlab{a}})\citenamefont {del Hougne}, \citenamefont
  {Imani}, \citenamefont {Sleasman}, \citenamefont {Gollub}, \citenamefont
  {Fink}, \citenamefont {Lerosey},\ and\ \citenamefont
  {Smith}}]{MotionDetector}%
  \BibitemOpen
  \bibfield  {author} {\bibinfo {author} {\bibfnamefont {P.}~\bibnamefont {del
  Hougne}}, \bibinfo {author} {\bibfnamefont {M.~F.}\ \bibnamefont {Imani}},
  \bibinfo {author} {\bibfnamefont {T.}~\bibnamefont {Sleasman}}, \bibinfo
  {author} {\bibfnamefont {J.~N.}\ \bibnamefont {Gollub}}, \bibinfo {author}
  {\bibfnamefont {M.}~\bibnamefont {Fink}}, \bibinfo {author} {\bibfnamefont
  {G.}~\bibnamefont {Lerosey}}, \ and\ \bibinfo {author} {\bibfnamefont
  {D.~R.}\ \bibnamefont {Smith}},\ }\bibfield  {title} {\enquote {\bibinfo
  {title} {Dynamic metasurface aperture as smart around-the-corner motion
  detector},}\ }\href@noop {} {\bibfield  {journal} {\bibinfo  {journal} {Sci.
  Rep.}\ }\textbf {\bibinfo {volume} {8}} (\bibinfo {year}
  {2018}{\natexlab{a}})}\BibitemShut {NoStop}%
\bibitem [{\citenamefont {del Hougne}\ \emph
  {et~al.}(2018{\natexlab{b}})\citenamefont {del Hougne}, \citenamefont
  {Imani}, \citenamefont {Fink}, \citenamefont {Smith},\ and\ \citenamefont
  {Lerosey}}]{Localization}%
  \BibitemOpen
  \bibfield  {author} {\bibinfo {author} {\bibfnamefont {P.}~\bibnamefont {del
  Hougne}}, \bibinfo {author} {\bibfnamefont {M.~F.}\ \bibnamefont {Imani}},
  \bibinfo {author} {\bibfnamefont {M.}~\bibnamefont {Fink}}, \bibinfo {author}
  {\bibfnamefont {D.~R.}\ \bibnamefont {Smith}}, \ and\ \bibinfo {author}
  {\bibfnamefont {G.}~\bibnamefont {Lerosey}},\ }\bibfield  {title} {\enquote
  {\bibinfo {title} {Precise localization of multiple noncooperative objects in
  a disordered cavity by wave front shaping},}\ }\href@noop {} {\bibfield
  {journal} {\bibinfo  {journal} {Phys. Rev. Lett.}\ }\textbf {\bibinfo
  {volume} {121}},\ \bibinfo {pages} {063901} (\bibinfo {year}
  {2018}{\natexlab{b}})}\BibitemShut {NoStop}%
\bibitem [{\citenamefont {Hong}\ \emph {et~al.}(2014)\citenamefont {Hong},
  \citenamefont {Mendez}, \citenamefont {Koch}, \citenamefont {Wall},\ and\
  \citenamefont {Anlage}}]{Anlage_NLlossyTR}%
  \BibitemOpen
  \bibfield  {author} {\bibinfo {author} {\bibfnamefont {S.~K.}\ \bibnamefont
  {Hong}}, \bibinfo {author} {\bibfnamefont {V.~M.}\ \bibnamefont {Mendez}},
  \bibinfo {author} {\bibfnamefont {T.}~\bibnamefont {Koch}}, \bibinfo {author}
  {\bibfnamefont {W.~S.}\ \bibnamefont {Wall}}, \ and\ \bibinfo {author}
  {\bibfnamefont {S.~M.}\ \bibnamefont {Anlage}},\ }\bibfield  {title}
  {\enquote {\bibinfo {title} {Nonlinear electromagnetic time reversal in an
  open semireverberant system},}\ }\href@noop {} {\bibfield  {journal}
  {\bibinfo  {journal} {Phys. Rev. Applied}\ }\textbf {\bibinfo {volume} {2}},\
  \bibinfo {pages} {044013} (\bibinfo {year} {2014})}\BibitemShut {NoStop}%
\bibitem [{\citenamefont {Smith}\ \emph {et~al.}(2017)\citenamefont {Smith},
  \citenamefont {Gowda}, \citenamefont {Yurduseven}, \citenamefont {Larouche},
  \citenamefont {Lipworth}, \citenamefont {Urzhumov},\ and\ \citenamefont
  {Reynolds}}]{WTP_Smith}%
  \BibitemOpen
  \bibfield  {author} {\bibinfo {author} {\bibfnamefont {D.~R.}\ \bibnamefont
  {Smith}}, \bibinfo {author} {\bibfnamefont {V.~R.}\ \bibnamefont {Gowda}},
  \bibinfo {author} {\bibfnamefont {O.}~\bibnamefont {Yurduseven}}, \bibinfo
  {author} {\bibfnamefont {S.}~\bibnamefont {Larouche}}, \bibinfo {author}
  {\bibfnamefont {G.}~\bibnamefont {Lipworth}}, \bibinfo {author}
  {\bibfnamefont {Y.}~\bibnamefont {Urzhumov}}, \ and\ \bibinfo {author}
  {\bibfnamefont {M.~S.}\ \bibnamefont {Reynolds}},\ }\bibfield  {title}
  {\enquote {\bibinfo {title} {An analysis of beamed wireless power transfer in
  the fresnel zone using a dynamic, metasurface aperture},}\ }\href@noop {}
  {\bibfield  {journal} {\bibinfo  {journal} {J. Appl. Phys.}\ }\textbf
  {\bibinfo {volume} {121}},\ \bibinfo {pages} {014901} (\bibinfo {year}
  {2017})}\BibitemShut {NoStop}%
\bibitem [{\citenamefont {del Hougne}\ \emph {et~al.}(2017)\citenamefont {del
  Hougne}, \citenamefont {Fink},\ and\ \citenamefont
  {Lerosey}}]{harvesting_arXiv}%
  \BibitemOpen
  \bibfield  {author} {\bibinfo {author} {\bibfnamefont {P.}~\bibnamefont {del
  Hougne}}, \bibinfo {author} {\bibfnamefont {M.}~\bibnamefont {Fink}}, \ and\
  \bibinfo {author} {\bibfnamefont {G.}~\bibnamefont {Lerosey}},\ }\bibfield
  {title} {\enquote {\bibinfo {title} {Shaping microwave fields using nonlinear
  unsolicited feedback: Application to enhance energy harvesting},}\
  }\href@noop {} {\bibfield  {journal} {\bibinfo  {journal} {Phys. Rev.
  Applied}\ }\textbf {\bibinfo {volume} {8}},\ \bibinfo {pages} {061001}
  (\bibinfo {year} {2017})}\BibitemShut {NoStop}%
\bibitem [{\citenamefont {Gowda}\ \emph {et~al.}(2018)\citenamefont {Gowda},
  \citenamefont {Imani}, \citenamefont {Sleasman}, \citenamefont {Yurduseven},\
  and\ \citenamefont {Smith}}]{vinay}%
  \BibitemOpen
  \bibfield  {author} {\bibinfo {author} {\bibfnamefont {V.~R.}\ \bibnamefont
  {Gowda}}, \bibinfo {author} {\bibfnamefont {M.~F.}\ \bibnamefont {Imani}},
  \bibinfo {author} {\bibfnamefont {T.}~\bibnamefont {Sleasman}}, \bibinfo
  {author} {\bibfnamefont {O.}~\bibnamefont {Yurduseven}}, \ and\ \bibinfo
  {author} {\bibfnamefont {D.~R.}\ \bibnamefont {Smith}},\ }\bibfield  {title}
  {\enquote {\bibinfo {title} {Focusing microwaves in the fresnel zone with a
  cavity-backed holographic metasurface},}\ }\href@noop {} {\bibfield
  {journal} {\bibinfo  {journal} {IEEE Access}\ }\textbf {\bibinfo {volume}
  {6}},\ \bibinfo {pages} {12815--12824} (\bibinfo {year} {2018})}\BibitemShut
  {NoStop}%
\bibitem [{\citenamefont {Hill}(2009)}]{hill_electromagnetic_2009}%
  \BibitemOpen
  \bibfield  {author} {\bibinfo {author} {\bibfnamefont {David~A}\ \bibnamefont
  {Hill}},\ }\href@noop {} {\emph {\bibinfo {title} {Electromagnetic Fields in
  Cavities: Deterministic and Statistical Theories}}},\ Vol.~\bibinfo {volume}
  {35}\ (\bibinfo  {publisher} {John Wiley \& Sons},\ \bibinfo {year}
  {2009})\BibitemShut {NoStop}%
\bibitem [{\citenamefont {Stein}\ \emph {et~al.}(1995)\citenamefont {Stein},
  \citenamefont {St{\"o}ckmann},\ and\ \citenamefont
  {Stoffregen}}]{stoffregen}%
  \BibitemOpen
  \bibfield  {author} {\bibinfo {author} {\bibfnamefont {J.}~\bibnamefont
  {Stein}}, \bibinfo {author} {\bibfnamefont {H.-J.}\ \bibnamefont
  {St{\"o}ckmann}}, \ and\ \bibinfo {author} {\bibfnamefont {U.}~\bibnamefont
  {Stoffregen}},\ }\bibfield  {title} {\enquote {\bibinfo {title} {Microwave
  studies of billiard green functions and propagators},}\ }\href@noop {}
  {\bibfield  {journal} {\bibinfo  {journal} {Phys. Rev. Lett.}\ }\textbf
  {\bibinfo {volume} {75}},\ \bibinfo {pages} {53} (\bibinfo {year}
  {1995})}\BibitemShut {NoStop}%
\bibitem [{\citenamefont {Barth{\'e}lemy}\ \emph {et~al.}(2005)\citenamefont
  {Barth{\'e}lemy}, \citenamefont {Legrand},\ and\ \citenamefont
  {Mortessagne}}]{SmatrixLegrand}%
  \BibitemOpen
  \bibfield  {author} {\bibinfo {author} {\bibfnamefont {J.}~\bibnamefont
  {Barth{\'e}lemy}}, \bibinfo {author} {\bibfnamefont {O.}~\bibnamefont
  {Legrand}}, \ and\ \bibinfo {author} {\bibfnamefont {F.}~\bibnamefont
  {Mortessagne}},\ }\bibfield  {title} {\enquote {\bibinfo {title} {Complete s
  matrix in a microwave cavity at room temperature},}\ }\href@noop {}
  {\bibfield  {journal} {\bibinfo  {journal} {Phys. Rev. E}\ }\textbf {\bibinfo
  {volume} {71}},\ \bibinfo {pages} {016205} (\bibinfo {year}
  {2005})}\BibitemShut {NoStop}%
\bibitem [{\citenamefont {Hemmady}\ \emph {et~al.}(2005)\citenamefont
  {Hemmady}, \citenamefont {Zheng}, \citenamefont {Antonsen~Jr}, \citenamefont
  {Ott},\ and\ \citenamefont {Anlage}}]{Anlage_S_Matrix}%
  \BibitemOpen
  \bibfield  {author} {\bibinfo {author} {\bibfnamefont {S.}~\bibnamefont
  {Hemmady}}, \bibinfo {author} {\bibfnamefont {X.}~\bibnamefont {Zheng}},
  \bibinfo {author} {\bibfnamefont {T.~M.}\ \bibnamefont {Antonsen~Jr}},
  \bibinfo {author} {\bibfnamefont {E.}~\bibnamefont {Ott}}, \ and\ \bibinfo
  {author} {\bibfnamefont {S.~M.}\ \bibnamefont {Anlage}},\ }\bibfield  {title}
  {\enquote {\bibinfo {title} {Universal statistics of the scattering
  coefficient of chaotic microwave cavities},}\ }\href@noop {} {\bibfield
  {journal} {\bibinfo  {journal} {Phys. Rev. E}\ }\textbf {\bibinfo {volume}
  {71}},\ \bibinfo {pages} {056215} (\bibinfo {year} {2005})}\BibitemShut
  {NoStop}%
\bibitem [{\citenamefont {Kuhl}\ \emph {et~al.}(2005)\citenamefont {Kuhl},
  \citenamefont {St{\"o}ckmann},\ and\ \citenamefont
  {Weaver}}]{kuhl2005classical}%
  \BibitemOpen
  \bibfield  {author} {\bibinfo {author} {\bibfnamefont {U.}~\bibnamefont
  {Kuhl}}, \bibinfo {author} {\bibfnamefont {H.-J.}\ \bibnamefont
  {St{\"o}ckmann}}, \ and\ \bibinfo {author} {\bibfnamefont {R.}~\bibnamefont
  {Weaver}},\ }\bibfield  {title} {\enquote {\bibinfo {title} {Classical wave
  experiments on chaotic scattering},}\ }\href@noop {} {\bibfield  {journal}
  {\bibinfo  {journal} {J. Phys. A}\ }\textbf {\bibinfo {volume} {38}},\
  \bibinfo {pages} {10433} (\bibinfo {year} {2005})}\BibitemShut {NoStop}%
\bibitem [{\citenamefont {Weyl}(1911)}]{WEYLoriginal}%
  \BibitemOpen
  \bibfield  {author} {\bibinfo {author} {\bibfnamefont {Hermann}\ \bibnamefont
  {Weyl}},\ }\bibfield  {title} {\enquote {\bibinfo {title} {{{\"U}ber die
  asymptotische Verteilung der Eigenwerte}},}\ }\href@noop {} {\bibfield
  {journal} {\bibinfo  {journal} {Nachrichten von der Gesellschaft der
  Wissenschaften zu G{\"o}ttingen, Mathematisch-Physikalische Klasse}\ }\textbf
  {\bibinfo {volume} {1911}},\ \bibinfo {pages} {110--117} (\bibinfo {year}
  {1911})}\BibitemShut {NoStop}%
\bibitem [{\citenamefont {Arendt}\ \emph {et~al.}(2009)\citenamefont {Arendt},
  \citenamefont {Nittka}, \citenamefont {Peter},\ and\ \citenamefont
  {Steiner}}]{WEYLbook}%
  \BibitemOpen
  \bibfield  {author} {\bibinfo {author} {\bibfnamefont {Wolfgang}\
  \bibnamefont {Arendt}}, \bibinfo {author} {\bibfnamefont {Robin}\
  \bibnamefont {Nittka}}, \bibinfo {author} {\bibfnamefont {Wolfgang}\
  \bibnamefont {Peter}}, \ and\ \bibinfo {author} {\bibfnamefont {Frank}\
  \bibnamefont {Steiner}},\ }\href@noop {} {\emph {\bibinfo {title} {Weyl’s
  law: Spectral Properties of the Laplacian in Mathematics and Physics}}}\
  (\bibinfo  {publisher} {Wiley-VCH Verlag GmbH \& Co. KGaA, Weinheim},\
  \bibinfo {year} {2009})\BibitemShut {NoStop}%
\bibitem [{\citenamefont {Dupr\'e}\ \emph {et~al.}(2015)\citenamefont
  {Dupr\'e}, \citenamefont {del Hougne}, \citenamefont {Fink}, \citenamefont
  {Lemoult},\ and\ \citenamefont {Lerosey}}]{publikation1}%
  \BibitemOpen
  \bibfield  {author} {\bibinfo {author} {\bibfnamefont {M.}~\bibnamefont
  {Dupr\'e}}, \bibinfo {author} {\bibfnamefont {P.}~\bibnamefont {del Hougne}},
  \bibinfo {author} {\bibfnamefont {M.}~\bibnamefont {Fink}}, \bibinfo {author}
  {\bibfnamefont {F.}~\bibnamefont {Lemoult}}, \ and\ \bibinfo {author}
  {\bibfnamefont {G.}~\bibnamefont {Lerosey}},\ }\bibfield  {title} {\enquote
  {\bibinfo {title} {Wave-field shaping in cavities: Waves trapped in a box
  with controllable boundaries},}\ }\href@noop {} {\bibfield  {journal}
  {\bibinfo  {journal} {Phys. Rev. Lett.}\ }\textbf {\bibinfo {volume} {115}},\
  \bibinfo {pages} {017701} (\bibinfo {year} {2015})}\BibitemShut {NoStop}%
\bibitem [{\citenamefont {Kaina}\ \emph
  {et~al.}(2014{\natexlab{a}})\citenamefont {Kaina}, \citenamefont {Dupr{\'e}},
  \citenamefont {Lerosey},\ and\ \citenamefont {Fink}}]{SMM_PoC}%
  \BibitemOpen
  \bibfield  {author} {\bibinfo {author} {\bibfnamefont {N.}~\bibnamefont
  {Kaina}}, \bibinfo {author} {\bibfnamefont {M.}~\bibnamefont {Dupr{\'e}}},
  \bibinfo {author} {\bibfnamefont {G.}~\bibnamefont {Lerosey}}, \ and\
  \bibinfo {author} {\bibfnamefont {M.}~\bibnamefont {Fink}},\ }\bibfield
  {title} {\enquote {\bibinfo {title} {Shaping complex microwave fields in
  reverberating media with binary tunable metasurfaces},}\ }\href@noop {}
  {\bibfield  {journal} {\bibinfo  {journal} {Sci. Rep.}\ }\textbf {\bibinfo
  {volume} {4}} (\bibinfo {year} {2014}{\natexlab{a}})}\BibitemShut {NoStop}%
\bibitem [{\citenamefont {Sievenpiper}\ \emph {et~al.}(1999)\citenamefont
  {Sievenpiper}, \citenamefont {Zhang}, \citenamefont {Broas}, \citenamefont
  {Alexopolous},\ and\ \citenamefont {Yablonovitch}}]{SievenpiperPMC}%
  \BibitemOpen
  \bibfield  {author} {\bibinfo {author} {\bibfnamefont {D.}~\bibnamefont
  {Sievenpiper}}, \bibinfo {author} {\bibfnamefont {L.}~\bibnamefont {Zhang}},
  \bibinfo {author} {\bibfnamefont {R.~F.~J.}\ \bibnamefont {Broas}}, \bibinfo
  {author} {\bibfnamefont {N.~G.}\ \bibnamefont {Alexopolous}}, \ and\ \bibinfo
  {author} {\bibfnamefont {E.}~\bibnamefont {Yablonovitch}},\ }\bibfield
  {title} {\enquote {\bibinfo {title} {High-impedance electromagnetic surfaces
  with a forbidden frequency band},}\ }\href@noop {} {\bibfield  {journal}
  {\bibinfo  {journal} {IEEE Trans. Microw. Theory Techn.}\ }\textbf {\bibinfo
  {volume} {47}},\ \bibinfo {pages} {2059--2074} (\bibinfo {year}
  {1999})}\BibitemShut {NoStop}%
\bibitem [{\citenamefont {Sievenpiper}\ \emph {et~al.}(2003)\citenamefont
  {Sievenpiper}, \citenamefont {Schaffner}, \citenamefont {Song}, \citenamefont
  {Loo},\ and\ \citenamefont {Tangonan}}]{SievenpiperTunable}%
  \BibitemOpen
  \bibfield  {author} {\bibinfo {author} {\bibfnamefont {D.~F.}\ \bibnamefont
  {Sievenpiper}}, \bibinfo {author} {\bibfnamefont {J.~H.}\ \bibnamefont
  {Schaffner}}, \bibinfo {author} {\bibfnamefont {H.~J.}\ \bibnamefont {Song}},
  \bibinfo {author} {\bibfnamefont {R.~Y.}\ \bibnamefont {Loo}}, \ and\
  \bibinfo {author} {\bibfnamefont {G.}~\bibnamefont {Tangonan}},\ }\bibfield
  {title} {\enquote {\bibinfo {title} {Two-dimensional beam steering using an
  electrically tunable impedance surface},}\ }\href@noop {} {\bibfield
  {journal} {\bibinfo  {journal} {‎IEEE Trans. Antennas Propag}\ }\textbf
  {\bibinfo {volume} {51}},\ \bibinfo {pages} {2713--2722} (\bibinfo {year}
  {2003})}\BibitemShut {NoStop}%
\bibitem [{\citenamefont {Sihvola}(2007)}]{SihvolaMetaOverview}%
  \BibitemOpen
  \bibfield  {author} {\bibinfo {author} {\bibfnamefont {A.}~\bibnamefont
  {Sihvola}},\ }\bibfield  {title} {\enquote {\bibinfo {title} {Metamaterials
  in electromagnetics},}\ }\href@noop {} {\bibfield  {journal} {\bibinfo
  {journal} {Metamaterials}\ }\textbf {\bibinfo {volume} {1}},\ \bibinfo
  {pages} {2--11} (\bibinfo {year} {2007})}\BibitemShut {NoStop}%
\bibitem [{\citenamefont {Kaina}\ \emph
  {et~al.}(2014{\natexlab{b}})\citenamefont {Kaina}, \citenamefont {Dupr\'{e}},
  \citenamefont {Fink},\ and\ \citenamefont {Lerosey}}]{SMM_design}%
  \BibitemOpen
  \bibfield  {author} {\bibinfo {author} {\bibfnamefont {N.}~\bibnamefont
  {Kaina}}, \bibinfo {author} {\bibfnamefont {M.}~\bibnamefont {Dupr\'{e}}},
  \bibinfo {author} {\bibfnamefont {M.}~\bibnamefont {Fink}}, \ and\ \bibinfo
  {author} {\bibfnamefont {G.}~\bibnamefont {Lerosey}},\ }\bibfield  {title}
  {\enquote {\bibinfo {title} {Hybridized resonances to design tunable binary
  phase metasurface unit cells},}\ }\href@noop {}
  {\bibfield  {journal} {\bibinfo  {journal} {Opt. Express}\ }\textbf {\bibinfo
  {volume} {22}},\ \bibinfo {pages} {18881--18888} (\bibinfo {year}
  {2014}{\natexlab{b}})}\BibitemShut {NoStop}%
\bibitem [{\citenamefont {Li}\ \emph {et~al.}(2018)\citenamefont {Li},
  \citenamefont {Singh},\ and\ \citenamefont
  {Sievenpiper}}]{SievenpiperReview}%
  \BibitemOpen
  \bibfield  {author} {\bibinfo {author} {\bibfnamefont {A.}~\bibnamefont
  {Li}}, \bibinfo {author} {\bibfnamefont {S.}~\bibnamefont {Singh}}, \ and\
  \bibinfo {author} {\bibfnamefont {D.}~\bibnamefont {Sievenpiper}},\
  }\bibfield  {title} {\enquote {\bibinfo {title} {Metasurfaces and their
  applications},}\ }\href@noop {} {\bibfield  {journal} {\bibinfo  {journal}
  {Nanophotonics}\ }\textbf {\bibinfo {volume} {7}},\ \bibinfo {pages}
  {989--1011} (\bibinfo {year} {2018})}\BibitemShut {NoStop}%
\bibitem [{\citenamefont {Anderson}(1958)}]{anderson}%
  \BibitemOpen
  \bibfield  {author} {\bibinfo {author} {\bibfnamefont {P.~W.}\ \bibnamefont
  {Anderson}},\ }\bibfield  {title} {\enquote {\bibinfo {title} {Absence of
  diffusion in certain random lattices},}\ }\href@noop {} {\bibfield  {journal}
  {\bibinfo  {journal} {Phys. Rev.}\ }\textbf {\bibinfo {volume} {109}},\
  \bibinfo {pages} {1492} (\bibinfo {year} {1958})}\BibitemShut {NoStop}%
\bibitem [{\citenamefont {Wang}\ and\ \citenamefont
  {Genack}(2011)}]{genack_transport_random_media}%
  \BibitemOpen
  \bibfield  {author} {\bibinfo {author} {\bibfnamefont {J.}~\bibnamefont
  {Wang}}\ and\ \bibinfo {author} {\bibfnamefont {A.~Z.}\ \bibnamefont
  {Genack}},\ }\bibfield  {title} {\enquote {\bibinfo {title} {Transport
  through modes in random media},}\ }\href@noop {} {\bibfield  {journal}
  {\bibinfo  {journal} {Nature}\ }\textbf {\bibinfo {volume} {471}},\ \bibinfo
  {pages} {345} (\bibinfo {year} {2011})}\BibitemShut {NoStop}%
\bibitem [{Note1()}]{Note1}%
  \BibitemOpen
  \bibinfo {note} {See Supplemental Material}\BibitemShut {NoStop}%
\bibitem [{\citenamefont {Derode}\ \emph {et~al.}(2001)\citenamefont {Derode},
  \citenamefont {Tourin},\ and\ \citenamefont {Fink}}]{derodePRE2}%
  \BibitemOpen
  \bibfield  {author} {\bibinfo {author} {\bibfnamefont {A.}~\bibnamefont
  {Derode}}, \bibinfo {author} {\bibfnamefont {A.}~\bibnamefont {Tourin}}, \
  and\ \bibinfo {author} {\bibfnamefont {M.}~\bibnamefont {Fink}},\ }\bibfield
  {title} {\enquote {\bibinfo {title} {Random multiple scattering of
  ultrasound. ii. is time reversal a self-averaging process?}}\ }\href@noop {}
  {\bibfield  {journal} {\bibinfo  {journal} {Phys. Rev. E}\ }\textbf {\bibinfo
  {volume} {64}},\ \bibinfo {pages} {036606} (\bibinfo {year}
  {2001})}\BibitemShut {NoStop}%
\bibitem [{\citenamefont {Sidiropoulos}\ \emph {et~al.}(2006)\citenamefont
  {Sidiropoulos}, \citenamefont {Davidson},\ and\ \citenamefont
  {Luo}}]{CSIbeamforming}%
  \BibitemOpen
  \bibfield  {author} {\bibinfo {author} {\bibfnamefont {N.~D.}\ \bibnamefont
  {Sidiropoulos}}, \bibinfo {author} {\bibfnamefont {T.~N.}\ \bibnamefont
  {Davidson}}, \ and\ \bibinfo {author} {\bibfnamefont {Z.-Q.}\ \bibnamefont
  {Luo}},\ }\bibfield  {title} {\enquote {\bibinfo {title} {Transmit
  beamforming for physical-layer multicasting},}\ }\href@noop {} {\bibfield
  {journal} {\bibinfo  {journal} {IEEE Trans. Signal Process}\ }\textbf
  {\bibinfo {volume} {54}},\ \bibinfo {pages} {2239--2251} (\bibinfo {year}
  {2006})}\BibitemShut {NoStop}%
\bibitem [{\citenamefont {Sen}\ \emph {et~al.}(2012)\citenamefont {Sen},
  \citenamefont {Radunovic}, \citenamefont {Choudhury},\ and\ \citenamefont
  {Minka}}]{souviksen}%
  \BibitemOpen
  \bibfield  {author} {\bibinfo {author} {\bibfnamefont {S.}~\bibnamefont
  {Sen}}, \bibinfo {author} {\bibfnamefont {B.}~\bibnamefont {Radunovic}},
  \bibinfo {author} {\bibfnamefont {R.~R.}\ \bibnamefont {Choudhury}}, \ and\
  \bibinfo {author} {\bibfnamefont {T.}~\bibnamefont {Minka}},\ }\bibfield
  {title} {\enquote {\bibinfo {title} {You are facing the mona lisa: spot
  localization using phy layer information},}\ }in\ \href@noop {} {\emph
  {\bibinfo {booktitle} {Proceedings of the 10th International Conference on
  Mobile Systems, Applications, and Services}}}\ (\bibinfo {organization}
  {ACM},\ \bibinfo {year} {2012})\ pp.\ \bibinfo {pages} {183--196}\BibitemShut
  {NoStop}%
\bibitem [{\citenamefont {Yang}\ \emph {et~al.}(2016)\citenamefont {Yang},
  \citenamefont {Cao}, \citenamefont {Yang}, \citenamefont {Gao}, \citenamefont
  {Xu}, \citenamefont {Li}, \citenamefont {Chen}, \citenamefont {Zhao},
  \citenamefont {Zheng},\ and\ \citenamefont {Li}}]{FPGA_metasurface}%
  \BibitemOpen
  \bibfield  {author} {\bibinfo {author} {\bibfnamefont {H.}~\bibnamefont
  {Yang}}, \bibinfo {author} {\bibfnamefont {X.}~\bibnamefont {Cao}}, \bibinfo
  {author} {\bibfnamefont {F.}~\bibnamefont {Yang}}, \bibinfo {author}
  {\bibfnamefont {J.}~\bibnamefont {Gao}}, \bibinfo {author} {\bibfnamefont
  {S.}~\bibnamefont {Xu}}, \bibinfo {author} {\bibfnamefont {M.}~\bibnamefont
  {Li}}, \bibinfo {author} {\bibfnamefont {X.}~\bibnamefont {Chen}}, \bibinfo
  {author} {\bibfnamefont {Y.}~\bibnamefont {Zhao}}, \bibinfo {author}
  {\bibfnamefont {Y.}~\bibnamefont {Zheng}}, \ and\ \bibinfo {author}
  {\bibfnamefont {S.}~\bibnamefont {Li}},\ }\bibfield  {title} {\enquote
  {\bibinfo {title} {A programmable metasurface with dynamic polarization,
  scattering and focusing control},}\ }\href@noop {} {\bibfield  {journal}
  {\bibinfo  {journal} {Sci. Rep.}\ }\textbf {\bibinfo {volume} {6}},\ \bibinfo
  {pages} {35692} (\bibinfo {year} {2016})}\BibitemShut {NoStop}%
\bibitem [{\citenamefont {del Hougne}\ \emph
  {et~al.}(2016{\natexlab{b}})\citenamefont {del Hougne}, \citenamefont
  {Rajaei}, \citenamefont {Daudet},\ and\ \citenamefont {Lerosey}}]{TM_RevMed}%
  \BibitemOpen
  \bibfield  {author} {\bibinfo {author} {\bibfnamefont {P.}~\bibnamefont {del
  Hougne}}, \bibinfo {author} {\bibfnamefont {B.}~\bibnamefont {Rajaei}},
  \bibinfo {author} {\bibfnamefont {L.}~\bibnamefont {Daudet}}, \ and\ \bibinfo
  {author} {\bibfnamefont {G.}~\bibnamefont {Lerosey}},\ }\bibfield  {title}
  {\enquote {\bibinfo {title} {Intensity-only measurement of partially
  uncontrollable transmission matrix: demonstration with wave-field shaping in
  a microwave cavity},}\ }\href@noop {} {\bibfield  {journal} {\bibinfo
  {journal} {Opt. Express}\ }\textbf {\bibinfo {volume} {24}},\ \bibinfo
  {pages} {18631--18641} (\bibinfo {year} {2016}{\natexlab{b}})}\BibitemShut
  {NoStop}%
\bibitem [{Note2()}]{Note2}%
  \BibitemOpen
  \bibinfo {note} {Interestingly, a similar scaling law proportional to
  $\protect \mathcal {N}^2$ is found for the number of components of a photonic
  circuit consisting of beam-splitters to implement a desired $\protect
  \mathcal {N}\times \protect \mathcal {N}$ linear operation \cite
  {reck1994experimental,MITonn}.}\BibitemShut {Stop}%
\bibitem [{\citenamefont {Dr{\'e}meau}\ \emph {et~al.}(2015)\citenamefont
  {Dr{\'e}meau}, \citenamefont {Liutkus}, \citenamefont {Martina},
  \citenamefont {Katz}, \citenamefont {Sch{\"u}lke}, \citenamefont {Krzakala},
  \citenamefont {Gigan},\ and\ \citenamefont {Daudet}}]{LD_binDMD}%
  \BibitemOpen
  \bibfield  {author} {\bibinfo {author} {\bibfnamefont {A.}~\bibnamefont
  {Dr{\'e}meau}}, \bibinfo {author} {\bibfnamefont {A.}~\bibnamefont
  {Liutkus}}, \bibinfo {author} {\bibfnamefont {D.}~\bibnamefont {Martina}},
  \bibinfo {author} {\bibfnamefont {O.}~\bibnamefont {Katz}}, \bibinfo {author}
  {\bibfnamefont {C.}~\bibnamefont {Sch{\"u}lke}}, \bibinfo {author}
  {\bibfnamefont {F.}~\bibnamefont {Krzakala}}, \bibinfo {author}
  {\bibfnamefont {S.}~\bibnamefont {Gigan}}, \ and\ \bibinfo {author}
  {\bibfnamefont {L.}~\bibnamefont {Daudet}},\ }\bibfield  {title} {\enquote
  {\bibinfo {title} {Reference-less measurement of the transmission matrix of a
  highly scattering material using a dmd and phase retrieval techniques},}\
  }\href@noop {} {\bibfield  {journal} {\bibinfo  {journal} {Opt. Express}\
  }\textbf {\bibinfo {volume} {23}},\ \bibinfo {pages} {11898--11911} (\bibinfo
  {year} {2015})}\BibitemShut {NoStop}%
\bibitem [{\citenamefont {Vellekoop}\ and\ \citenamefont
  {Mosk}(2008)}]{moskWSalgo}%
  \BibitemOpen
  \bibfield  {author} {\bibinfo {author} {\bibfnamefont {I.~M.}\ \bibnamefont
  {Vellekoop}}\ and\ \bibinfo {author} {\bibfnamefont {A.~P.}\ \bibnamefont
  {Mosk}},\ }\bibfield  {title} {\enquote {\bibinfo {title} {Phase control
  algorithms for focusing light through turbid media},}\ }\href@noop {}
  {\bibfield  {journal} {\bibinfo  {journal} {Opt. Commun.}\ }\textbf {\bibinfo
  {volume} {281}},\ \bibinfo {pages} {3071--3080} (\bibinfo {year}
  {2008})}\BibitemShut {NoStop}%
\bibitem [{\citenamefont {Bertsimas}\ and\ \citenamefont
  {Nohadani}(2010)}]{SimulAnnneal}%
  \BibitemOpen
  \bibfield  {author} {\bibinfo {author} {\bibfnamefont {D.}~\bibnamefont
  {Bertsimas}}\ and\ \bibinfo {author} {\bibfnamefont {O.}~\bibnamefont
  {Nohadani}},\ }\bibfield  {title} {\enquote {\bibinfo {title} {Robust
  optimization with simulated annealing},}\ }\href@noop {} {\bibfield
  {journal} {\bibinfo  {journal} {J. Glob. Optim.}\ }\textbf {\bibinfo {volume}
  {48}},\ \bibinfo {pages} {323--334} (\bibinfo {year} {2010})}\BibitemShut
  {NoStop}%
\bibitem [{\citenamefont {Peurifoy}\ \emph {et~al.}(2018)\citenamefont
  {Peurifoy}, \citenamefont {Shen}, \citenamefont {Jing}, \citenamefont {Yang},
  \citenamefont {Cano-Renteria}, \citenamefont {DeLacy}, \citenamefont
  {Joannopoulos}, \citenamefont {Tegmark},\ and\ \citenamefont
  {Solja{\v{c}}i{\'c}}}]{SciAdvANNinvDesing}%
  \BibitemOpen
  \bibfield  {author} {\bibinfo {author} {\bibfnamefont {J.}~\bibnamefont
  {Peurifoy}}, \bibinfo {author} {\bibfnamefont {Y.}~\bibnamefont {Shen}},
  \bibinfo {author} {\bibfnamefont {L.}~\bibnamefont {Jing}}, \bibinfo {author}
  {\bibfnamefont {Y.}~\bibnamefont {Yang}}, \bibinfo {author} {\bibfnamefont
  {F.}~\bibnamefont {Cano-Renteria}}, \bibinfo {author} {\bibfnamefont {B.~G.}\
  \bibnamefont {DeLacy}}, \bibinfo {author} {\bibfnamefont {J.~D.}\
  \bibnamefont {Joannopoulos}}, \bibinfo {author} {\bibfnamefont
  {M.}~\bibnamefont {Tegmark}}, \ and\ \bibinfo {author} {\bibfnamefont
  {M.}~\bibnamefont {Solja{\v{c}}i{\'c}}},\ }\bibfield  {title} {\enquote
  {\bibinfo {title} {Nanophotonic particle simulation and inverse design using
  artificial neural networks},}\ }\href@noop {} {\bibfield  {journal} {\bibinfo
   {journal} {Sci. Adv.}\ }\textbf {\bibinfo {volume} {4}},\ \bibinfo {pages}
  {eaar4206} (\bibinfo {year} {2018})}\BibitemShut {NoStop}%
\bibitem [{\citenamefont {van Putten}\ \emph {et~al.}(2011)\citenamefont {van
  Putten}, \citenamefont {Akbulut}, \citenamefont {Bertolotti}, \citenamefont
  {Vos}, \citenamefont {Lagendijk},\ and\ \citenamefont
  {Mosk}}]{Mosk_vis_subwavelength_foc_byWS}%
  \BibitemOpen
  \bibfield  {author} {\bibinfo {author} {\bibfnamefont {E.~G.}\ \bibnamefont
  {van Putten}}, \bibinfo {author} {\bibfnamefont {D.}~\bibnamefont {Akbulut}},
  \bibinfo {author} {\bibfnamefont {J.}~\bibnamefont {Bertolotti}}, \bibinfo
  {author} {\bibfnamefont {W.~L.}\ \bibnamefont {Vos}}, \bibinfo {author}
  {\bibfnamefont {A.}~\bibnamefont {Lagendijk}}, \ and\ \bibinfo {author}
  {\bibfnamefont {A.~P.}\ \bibnamefont {Mosk}},\ }\bibfield  {title} {\enquote
  {\bibinfo {title} {Scattering lens resolves sub-100 nm structures with
  visible light},}\ }\href@noop {} {\bibfield
   {journal} {\bibinfo  {journal} {Phys. Rev. Lett.}\ }\textbf {\bibinfo
  {volume} {106}},\ \bibinfo {pages} {193905} (\bibinfo {year}
  {2011})}\BibitemShut {NoStop}%
\bibitem [{\citenamefont {Sze}\ \emph {et~al.}(2017)\citenamefont {Sze},
  \citenamefont {Chen}, \citenamefont {Yang},\ and\ \citenamefont
  {Emer}}]{EfficientDNN}%
  \BibitemOpen
  \bibfield  {author} {\bibinfo {author} {\bibfnamefont {V.}~\bibnamefont
  {Sze}}, \bibinfo {author} {\bibfnamefont {Y.-H.}\ \bibnamefont {Chen}},
  \bibinfo {author} {\bibfnamefont {T.-J.}\ \bibnamefont {Yang}}, \ and\
  \bibinfo {author} {\bibfnamefont {J.~S.}\ \bibnamefont {Emer}},\ }\bibfield
  {title} {\enquote {\bibinfo {title} {Efficient processing of deep neural
  networks: A tutorial and survey},}\ }\href@noop {} {\bibfield  {journal}
  {\bibinfo  {journal} {Proc. IEEE}\ }\textbf {\bibinfo {volume} {105}},\
  \bibinfo {pages} {2295--2329} (\bibinfo {year} {2017})}\BibitemShut {NoStop}%
\bibitem [{\citenamefont {Jia}\ \emph {et~al.}(2014)\citenamefont {Jia},
  \citenamefont {Shelhamer}, \citenamefont {Donahue}, \citenamefont {Karayev},
  \citenamefont {Long}, \citenamefont {Girshick}, \citenamefont {Guadarrama},\
  and\ \citenamefont {Darrell}}]{caffe}%
  \BibitemOpen
  \bibfield  {author} {\bibinfo {author} {\bibfnamefont {Y.}~\bibnamefont
  {Jia}}, \bibinfo {author} {\bibfnamefont {E.}~\bibnamefont {Shelhamer}},
  \bibinfo {author} {\bibfnamefont {J.}~\bibnamefont {Donahue}}, \bibinfo
  {author} {\bibfnamefont {S.}~\bibnamefont {Karayev}}, \bibinfo {author}
  {\bibfnamefont {J.}~\bibnamefont {Long}}, \bibinfo {author} {\bibfnamefont
  {R.}~\bibnamefont {Girshick}}, \bibinfo {author} {\bibfnamefont
  {S.}~\bibnamefont {Guadarrama}}, \ and\ \bibinfo {author} {\bibfnamefont
  {T.}~\bibnamefont {Darrell}},\ }\bibfield  {title} {\enquote {\bibinfo
  {title} {Caffe: Convolutional architecture for fast feature embedding},}\
  }in\ \href@noop {} {\emph {\bibinfo {booktitle} {Proceedings of the 22nd ACM
  international conference on Multimedia}}}\ (\bibinfo {organization} {ACM},\
  \bibinfo {year} {2014})\ pp.\ \bibinfo {pages} {675--678}\BibitemShut
  {NoStop}%
\bibitem [{\citenamefont {Gros}\ \emph {et~al.}(2018)\citenamefont {Gros},
  \citenamefont {del Hougne},\ and\ \citenamefont {Lerosey}}]{CEM2018}%
  \BibitemOpen
  \bibfield  {author} {\bibinfo {author} {\bibfnamefont {J.-B.}\ \bibnamefont
  {Gros}}, \bibinfo {author} {\bibfnamefont {P.}~\bibnamefont {del Hougne}}, \
  and\ \bibinfo {author} {\bibfnamefont {G.}~\bibnamefont {Lerosey}},\
  }\href@noop {} {\enquote {\bibinfo {title} {Les cavit\'{e}s micro-ondes
  reconfigurables: Une nouvelle approche pour fabriquer des chambres
  r\'{e}verb\'{e}rantes chaotiques},}\ } (\bibinfo {year} {2018}),\ \bibinfo
  {note} {19\`{e}me Colloque International \& Exposition sur la
  Compatibilit\'{e} ElectroMagn\'{e}tique}\BibitemShut {NoStop}%
\bibitem [{\citenamefont {Serra}\ \emph {et~al.}(2016)\citenamefont {Serra},
  \citenamefont {Marvin}, \citenamefont {Moglie}, \citenamefont {Primiani},
  \citenamefont {Cozza}, \citenamefont {Arnaut}, \citenamefont {Huang},
  \citenamefont {Hatfield}, \citenamefont {Klingler},\ and\ \citenamefont
  {Leferink}}]{ramiroEMC}%
  \BibitemOpen
  \bibfield  {author} {\bibinfo {author} {\bibfnamefont {R.}~\bibnamefont
  {Serra}}, \bibinfo {author} {\bibfnamefont {A.~C.}\ \bibnamefont {Marvin}},
  \bibinfo {author} {\bibfnamefont {F.}~\bibnamefont {Moglie}}, \bibinfo
  {author} {\bibfnamefont {V.~M.}\ \bibnamefont {Primiani}}, \bibinfo {author}
  {\bibfnamefont {A.}~\bibnamefont {Cozza}}, \bibinfo {author} {\bibfnamefont
  {L.~R.}\ \bibnamefont {Arnaut}}, \bibinfo {author} {\bibfnamefont
  {Y.}~\bibnamefont {Huang}}, \bibinfo {author} {\bibfnamefont {M.~O.}\
  \bibnamefont {Hatfield}}, \bibinfo {author} {\bibfnamefont {M.}~\bibnamefont
  {Klingler}}, \ and\ \bibinfo {author} {\bibfnamefont {F.}~\bibnamefont
  {Leferink}},\ }\bibfield  {title} {\enquote {\bibinfo {title} {Reverberation
  chambers {\`a} la carte: An overview of the different mode-stirring
  techniques},}\ }\href@noop {} {\bibfield  {journal} {\bibinfo  {journal}
  {IEEE Electrmagn. Compat}\ }\textbf {\bibinfo {volume} {6}},\ \bibinfo
  {pages} {63--78} (\bibinfo {year} {2016})}\BibitemShut {NoStop}%
\bibitem [{\citenamefont {Horowitz}(2014)}]{horowitz20141}%
  \BibitemOpen
  \bibfield  {author} {\bibinfo {author} {\bibfnamefont {M.}~\bibnamefont
  {Horowitz}},\ }\bibfield  {title} {\enquote {\bibinfo {title} {1.1
  computing's energy problem (and what we can do about it)},}\ }in\ \href@noop
  {} {\emph {\bibinfo {booktitle} {Solid-State Circuits Conference Digest of
  Technical Papers (ISSCC), 2014 IEEE International}}}\ (\bibinfo
  {organization} {IEEE},\ \bibinfo {year} {2014})\ pp.\ \bibinfo {pages}
  {10--14}\BibitemShut {NoStop}%
\bibitem [{\citenamefont {Pratt}\ and\ \citenamefont
  {Jennings}(1996)}]{TransferLearningPreciseCitation}%
  \BibitemOpen
  \bibfield  {author} {\bibinfo {author} {\bibfnamefont {L.}~\bibnamefont
  {Pratt}}\ and\ \bibinfo {author} {\bibfnamefont {B.}~\bibnamefont
  {Jennings}},\ }\bibfield  {title} {\enquote {\bibinfo {title} {A survey of
  connectionist network reuse through transfer},}\ }in\ \href@noop {} {\emph
  {\bibinfo {booktitle} {Learning to learn}}}\ (\bibinfo  {publisher}
  {Springer},\ \bibinfo {year} {1996})\ pp.\ \bibinfo {pages}
  {19--43}\BibitemShut {NoStop}%
\bibitem [{\citenamefont {Ambichl}\ \emph {et~al.}(2017)\citenamefont
  {Ambichl}, \citenamefont {Xiong}, \citenamefont {Bromberg}, \citenamefont
  {Redding}, \citenamefont {Cao},\ and\ \citenamefont {Rotter}}]{RotterPRX}%
  \BibitemOpen
  \bibfield  {author} {\bibinfo {author} {\bibfnamefont {P.}~\bibnamefont
  {Ambichl}}, \bibinfo {author} {\bibfnamefont {W.}~\bibnamefont {Xiong}},
  \bibinfo {author} {\bibfnamefont {Y.}~\bibnamefont {Bromberg}}, \bibinfo
  {author} {\bibfnamefont {B.}~\bibnamefont {Redding}}, \bibinfo {author}
  {\bibfnamefont {H.}~\bibnamefont {Cao}}, \ and\ \bibinfo {author}
  {\bibfnamefont {S.}~\bibnamefont {Rotter}},\ }\bibfield  {title} {\enquote
  {\bibinfo {title} {Super-and anti-principal-modes in multimode waveguides},}\
  }\href@noop {} {\bibfield  {journal} {\bibinfo  {journal} {Phys. Rev. X}\
  }\textbf {\bibinfo {volume} {7}},\ \bibinfo {pages} {041053} (\bibinfo {year}
  {2017})}\BibitemShut {NoStop}%
\end{thebibliography}


%

\renewcommand{\thefigure}{S\arabic{figure}}
\renewcommand{\theequation}{S\arabic{equation}}
\setcounter{figure}{0}
\setcounter{equation}{0}

\clearpage
\section*{Supplemental Material}
For the interested reader, here we provide numerous additional details that complement the manuscript and provide further illustrations. This document is organized as follows:

A. Experimental Setup.

B. Long-Range Correlations in the Impact Matrix.

C. Ensemble Averaging over Realizations.

D. Time-Sequential WBAC Scheme.

E. Energy Efficiency.

F. Notation.

\subsection{Experimental Setup}

\textit{Quality Factor and Correlation Frequency.} --- Our metallic cavity's geometry is clearly irregular due to a few irregularly shaped metallic objects glued to the walls as well as the presence of the large mode-stirrer (see Fig.~3 in the main text). Pieces of electromagnetic absorbers ($10\times 10\times 5\ \mathrm{cm}^3$) are isotropically distributed across the cavity's walls, lowering its quality factor from $1023$ to $179$. This is on the order of the quality factor of $234$ that we have measured in an office room in our laboratory. We estimate $Q$ as $\pi f_0/\mu$, where $\mu$ is the exponential decay constant extracted from the envelope of the inverse Fourier transform of the experimentally measured broadband Green's function, averaged over $30$ mode-stirrer positions.
Our cavity's correlation frequency is thus $\Delta f_{\mathrm{corr}} = f_0 / Q \approx 15 \ \mathrm{MHz}$. Consequently we choose five working frequencies (within the reflect-array's operating band) that are separated by at least $\Delta f_{\mathrm{corr}}$ in order for them to constitute independent realizations.

\textit{Tunable Metasurface Reflect-Array.} --- The metasurface reflect-array consists of $n=88$ functional elements ($8$ of its $96$ elements are broken) whose design is essentially the one presented in Ref.~\cite{SMM_design}. However, each element has independent control over the two polarizations of the electromagnetic field, rather than acting only on one polarization as in Ref.~\cite{SMM_design}. We synchronize each element's horizontal and vertical polarization configuration to simplify our experiment. Each of our monopole antennas picks up the field polarized along its axis; since polarizations are mixed in a chaotic cavity we have $n=88$ controllable elements whose average impact on the wave field is twice as much as it would be if they controlled only one polarization. Thus, we effectively control $p\approx 2\times(n/3) = 59$ modes, based on the result from Ref.~\cite{publikation1} that a metasurface consisting of $n$ phase-binary pixels acting on a single polarization controls about $n/3$ modes.

\subsection{Long-Range Correlations in the Impact Matrix}

The impact matrix (IM) formalism links the vector $\mathrm{V}$ describing the configuration of the reflect-array elements to the measured transmissions $\mathrm{Y}$ via $\mathrm{
Y}=\mathbf{H}\mathrm{V}$ \cite{TM_RevMed}. Here we illustrate the long-range correlations between IM entries in our experimental setup. As the wave reverberates inside the cavity, it revisits multiple times the reflect-array such that the impact of reflect-array element $i$ on the transmission between the pair of antennas indexed $j$, that is the entry $H_{i,j}$ of the impact matrix $\mathbf{H}$, depends to some extent on the configuration of the reflect-array elements other than $i$, too.  

\begin{figure}[bt]
	\begin{center}
\includegraphics [width=\columnwidth] {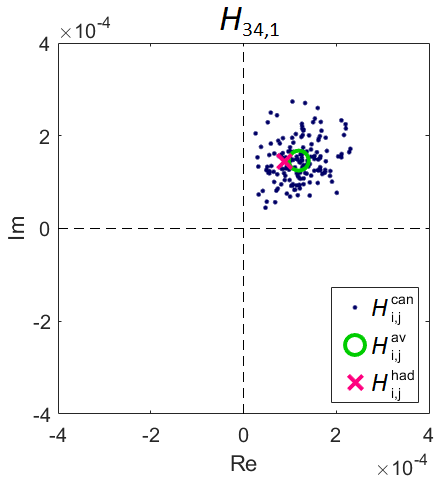}
	\caption{Visualization of impact matrix (IM) long-range correlations. For a given cavity geometry, we show measurements of the IM entry $H_{34,1}$ linking the metasurface reflect-array element $34$ to the transmission measurement indexed $1$. We illustrate the cloud of values obtained from measurements in the canonical basis as outlined above (blue dots), their average (green circle) as well as the value measured in the Hadamard basis (violet cross). We used the latter throughout the main text. The measurements in the canonical basis are taken for random configurations of the elements other than the one under consideration, clearly showing that $H_{i,j}$ significantly depends on how the elements $k \neq i$ are configured.} 
	\label{fig_H_clouds}
	\end{center}
\end{figure}

To visualize the long-range correlations, we measure $H_{i,j}$ in our experiment for a given realization with different configurations of the reflect-array elements other than $i$. For notational ease, let $H_{i,j}^{had}$ be the IM entry measured in the Hadamard basis, as outlined in Section~III.D of the main text. To now obtain many measurements of $H_{i,j}$, we work in the usual, canonical basis. We define $\mathrm{V}^+$ and $\mathrm{V}^-$ as being equal to the same random phase-binary configuration $\mathrm{V}$, except for the $i$th entry that is fixed to be $+1$ or $-1$ for $\mathrm{V}^+$ and $\mathrm{V}^-$, respectively. By measuring the corresponding transmissions $\mathrm{Y}^+$ and $\mathrm{Y}^-$, we can obtain an estimate of $H_{i,j}$ as follows: $H_{i,j}^{can} = ( Y_j^+ - Y_j^- )/2$. We repeat this $160$ times with different random vectors $\mathrm{V}$, yielding $160$ estimates of $H_{i,j}^{can}$. From those, we can further define an average over the $160$ estimates: $H_{i,j}^{av} = \langle  H_{i,j}^{can} \rangle$.
In Fig.~\ref{fig_H_clouds} we display 
in the complex plane the experimentally measured cloud of $160$ points for $H_{i,j}^{can}$, their average $H_{i,j}^{av}$ as well as $H_{i,j}^{had}$. 

The long-range correlations are clearly visualized by the non-negligible radius of the blue cloud. The numerical identification of the appropriate configuration for a given wave front segment utilizes one value of $H_{i,j}$. Yet, the actual value in the computation phase will depend on how the reflect-array is configured then. Clearly, using $H_{i,j}^{av}$ in the numerical optimization is the best choice. It is, however, very cumbersome to measure. The measurement in the Hadamard basis that we opt for requires only a single measurement per IM entry and has an averaging effect because it employs all reflect-array elements. It therefore seems to be a good trade-off between measurement duration and accuracy. The averaging over realizations is inevitable in any case in the cavity implementation of our scheme, but the further from $H_{i,j}^{av}$ the value used in the numerical optimization is, the more likely it is that more realizations are needed to converge to the exact computation results.

Finally, we note that in Ref.~\cite{TM_RevMed} a first-order approximation of an IM in a similar system (low-$Q$ irregularly-shaped microwave cavity) was measured and successfully used for focusing without ensemble averaging over realizations. This worked because said focusing experiment only required a phase-binary decision in order to align the contribution from each pixel of the WFS device with the overall desired direction in the complex plane. As seen in Fig.~\ref{fig_H_clouds}
, in a cavity with high modal density and $N>p$ the cloud of possible $H_{i,j}$ values usually remains within one quadrant, here the upper right one. The focusing decision thus only requires a choice as to whether we want some value in the upper right (element configuration $r=+1$) or lower left (element configuration $r=-1$) quadrant to maximize the length of the resultant. In the present work, the optimization constraints are significantly more challenging to satisfy: a specific amplitude \textit{and} phase has to be achieved with the tamed walk, \textit{simultaneously} at several observation points.

\subsection{Ensemble Averaging over Realizations}

\begin{figure}[t]
	\begin{center}
\includegraphics [width=\columnwidth] {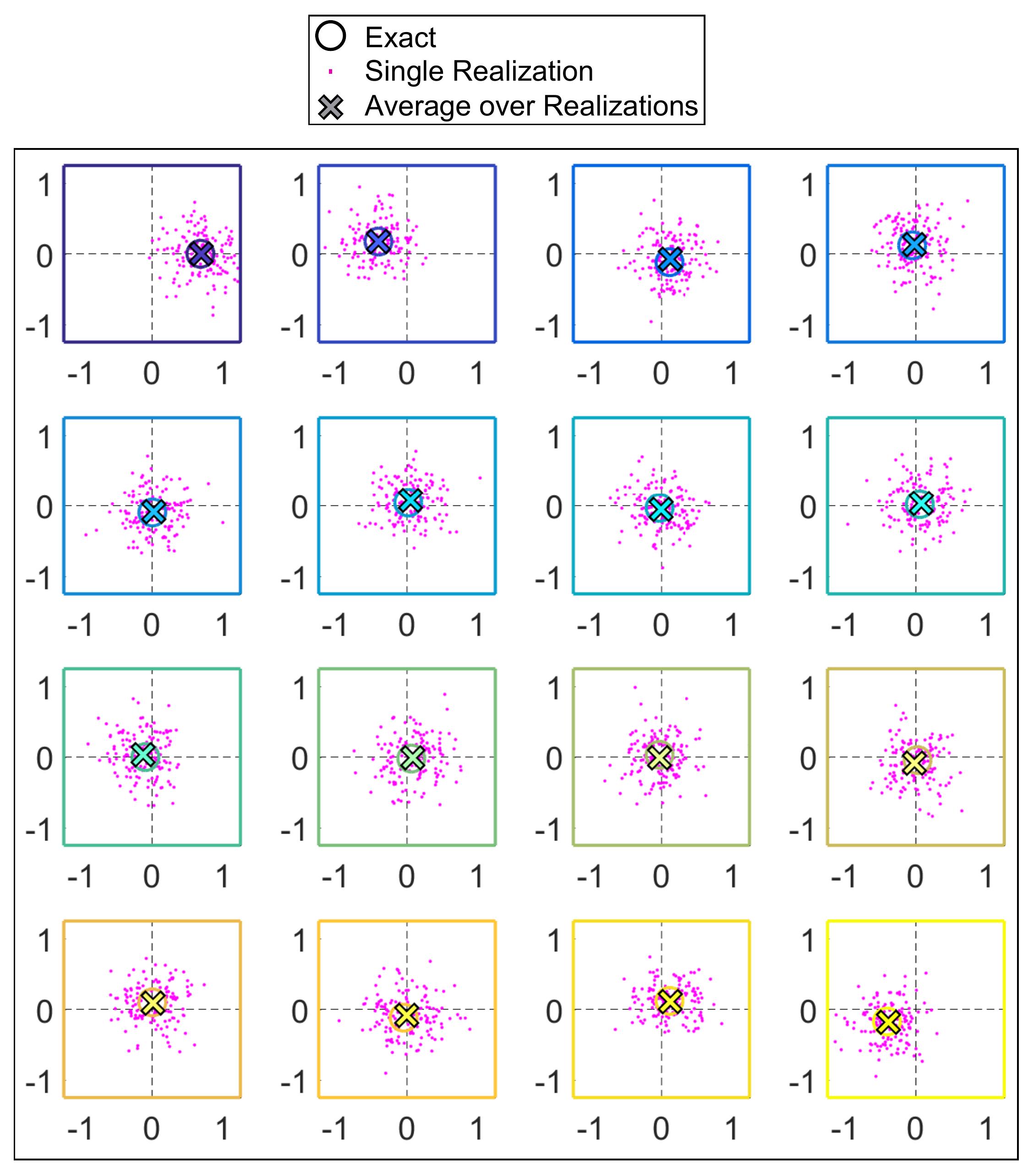}
	\caption{Visualization of the outcomes of the $150$ individual realizations, as well as their average that is displayed in Fig.~5(a) in the main text.} 
	\label{figAEC}
	\end{center}
\end{figure}

In Fig.~\ref{figAEC}, for the experiment of computing the Fourier transform of the Eiffel Tower's silhouette (see main text Fig.~5(a)), we display for each output point the outcomes of the $150$ individual realizations ($30$ mode-stirrer positions, $5$  independent working frequencies), as well as their average and the exact value that are shown in Fig.~5(a).

\subsection{Time-Sequential WBAC Scheme}

In this section, we first visually illustrate the principle of sequential WBAC explained in the main text. Then, we provide the $16\times 16$ complex-valued discrete Fourier transform operation $\mathbf{G}$ that we implemented, corresponding to Fig.~5 in the main text. Finally, we show some further computation examples obtained with said operation that we did not include in the main 
text for conciseness.

\subsubsection{Visual Illustration of Sequential WBAC}

\begin{figure}[h]
	\begin{center}
\includegraphics [width=\columnwidth] {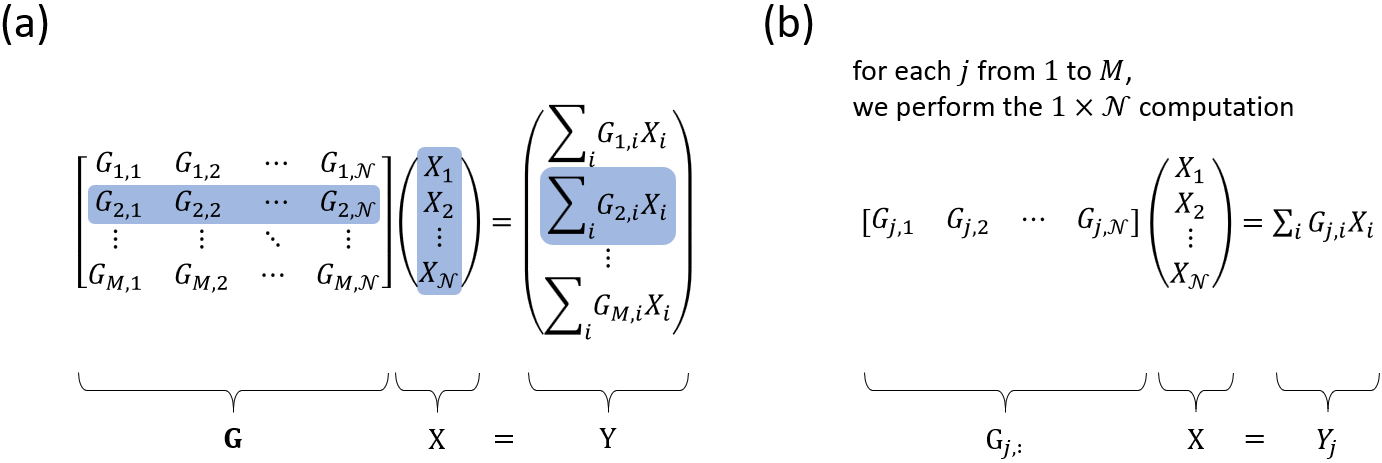}
	\caption{Visualization of the sequential WBAC scheme introduced in the main text. The output vector entries are independent from each other (a) and may be computed one after the other (b).} 
	\label{fig_VisSeq}
	\end{center}
\end{figure}

Here, we provide a visualization of how we can achieve a much bigger effective computation unit than our fixed-size physical system. Consider the general matrix formalism for an operation $\mathbf{G}$ as shown in Fig.~\ref{fig_VisSeq}(a). For an $M\times \mathcal{N}$ operation $\mathbf{G}$, each output entry in $\mathrm{Y}$ is in fact the product of the corresponding row of $\mathbf{G}$, itself a vector, and the input vector $\mathrm{X}$. We can thus separately compute each output entry in a $1\times \mathcal{N}$ operation. 

Since in the time-sequential scheme the resultants of the complex walks (see Fig.~2 in the main text) have to be tamed only for one observation point at any given time, the optimization constraints are considerably relaxed (but still harder than simple focusing, see Section~III.F of the main text) such that it becomes possible to implement larger operations with more inputs and outputs than in a single-shot computation scheme. The procedure is thus essentially looped over each of the output vector entries, as schematically shown in Fig.~\ref{fig_VisSeq}(b). Moreover, this means that we only need one transmission measurement for the sequential version. Note that one may in principle also compute more than one of the output vector's entries per loop to speed up the computation time.

\subsubsection{The implemented $16\times 16$ complex-valued operation $\mathbf{G}$}

The complex-valued $16\times 16$ discrete Fourier transform operation implemented to produce Fig.~5 of the main text is

\begin{figure}[h]
	\begin{center}
\includegraphics [width=7 cm] {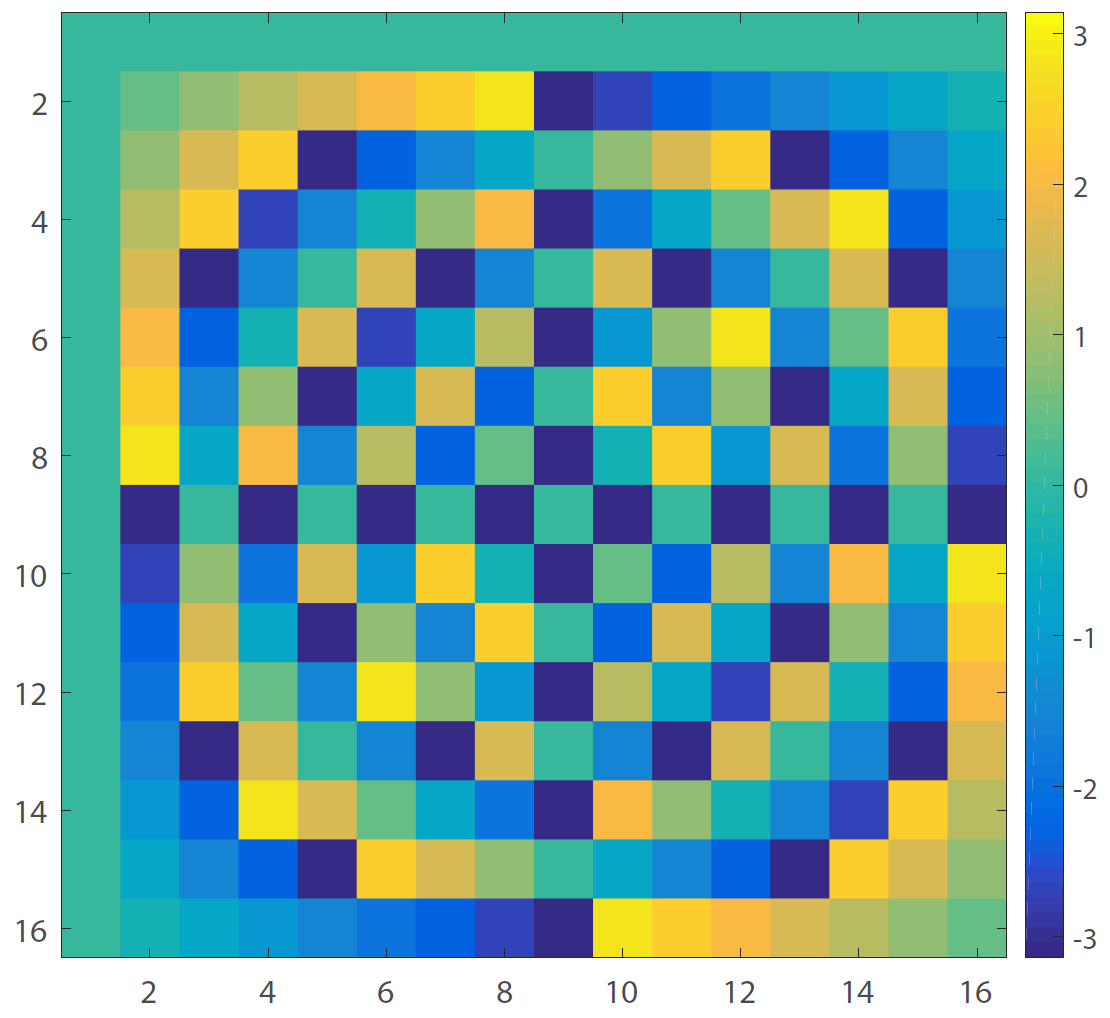}
	\caption{Phase of $\mathbf{G}$ (see Eq.~\ref{eqnG16}); the magnitude is unity for all entries.} 
	\label{fig_G16}
	\end{center}
\end{figure}

\begin{equation}\label{eqnG16}
 \mathbf{G} =
  \begin{bmatrix}
    1 & 1 & 1 & 1 & \cdots & 1 \\
    1 & \omega_z & \omega_z^2 & \omega_z^3 & \cdots & \omega_z^{(Z-1)}   \\
    1 & \omega_z^2 & \omega_z^4 & \omega_z^6 & \cdots & \omega_z^{2(Z-1)}   \\
    1 & \omega_z^3 & \omega_z^6 & \omega_z^9 & \cdots & \omega_z^{3(Z-1)}   \\
    \vdots & \vdots & \vdots & \vdots & \ddots & \vdots   \\
    1 & \omega_z^{(Z-1)} & \omega_z^{2(Z-1)} & \omega_z^{3(Z-1)}  & \cdots & \omega_z^{(Z-1)(Z-1)} 
  \end{bmatrix},
\end{equation}

\bigskip

\noindent where $\omega_z = \mathrm{exp}\left( \frac{2\pi i}{2^z} \right)$ is the $z$th root of unity and we have $z=Z=16$. The magnitudes of all entries of $\mathbf{G}$ are unity, their phases are illustrated in Fig.~\ref{fig_G16}.

\subsubsection{Further sample sequential computation results}

In Fig.~\ref{fig_further_examples} we provide further examples of $16\times 16$ computations that we carried out with $\mathbf{G}$ as defined in Eq.~\ref{eqnG16} to complement the main text where two examples using high-resolution input vectors $\mathrm{X}$ that correspond to the silhouettes of famous Parisian monuments were shown. In the following, we show results corresponding to some specific input vectors that are interesting mathematically (vectors with periodic entries, for instance). We display the results both in the complex plane (left) and in terms of the corresponding intensities (right). The results were obtained by ensemble averaging over $150$ realizations, like the results in the main text.

\subsection{Energy Efficiency}

For our cavity implementation, three factors potentially dominating the energy consumption come to mind: (i) generating the wave field, (ii) controlling the metasurface configuration with a micro-controller, and (iii) the feeding circuit for the tunable elements in the metasurface. Given the fantastic signal-to-noise ratio in the shielded cavity, low-power operation with emission of the continuous wave at $-20\ \mathrm{dB}$ or lower is certainly realistic, implying that less than $10\ \mathrm{\mu W}$ are consumed for (i). The micro-controllers do not consume more than a few $\mathrm{mW}$ per hundreds of channels \cite{SMM_PoC}. The power consumed my the metasurface feeding circuit used in Ref.~\cite{SMM_PoC} is $25\ \mathrm{\mu W / pixel}$, and thus, for instance, $12.5\ \mathrm{mW}$ for a system with $500$ pixels. Although the current through the PIN diodes could be further reduced to lower the feeding circuit's power consumption at the price of more wave attenuation on the metasurface, (iii) appears to be the dominating factor when a large number of pixels is used.

\subsection{Notation}

For reference and clarity's sake, we summarize the notation used throughout our main text and this supplemental material to denote matrices, vectors and numbers, and to refer to specific entries.

\begin{itemize}
\item Matrices are typed in bold, non-italic face, e.g. $\mathbf{H}$, $\mathbf{G}$.
\item Vectors are typed in non-italic face, e.g. $\mathrm{X}$, $\mathrm{Y}$, $\mathrm{V}$.
\item Single numbers are typed in italic face, e.g. $\alpha$, $\gamma$.
\end{itemize}

To refer to a specific vector entry, the subscript will provide the relevant index information, e.g. $Y_j$ is the $j$th entry of vector $\mathrm{Y}$.

To refer to a specific matrix entry, the subscript will provide row and column information separated by a comma, e.g. $H_{i,j}$ is the entry in the $i$th row and $j$th column of $\mathbf{H}$ (and hence a single number).

To refer to multiple specific entries of a vector, the subscript will provide the relevant index information, e.g. $\mathrm{V}_D$ are the entries of vector $\mathrm{V}$ that form part of the group of WFS device pixels labeled $D$. Note that $\mathrm{V}_D$ is itself a vector, hence it is typed in non-italic face.

To refer to multiple specific entries of a matrix, the subscript will provide the relevant information, e.g. $\mathrm{H}_{D,j}$ are the entries found in the rows forming part of group $D$ and column $j$. Note that $\mathrm{H}_{D,j}$ is itself a vector, hence it is typed in non-italic face. $\mathrm{G}_{j,:}$ refers to the entire $j$th row of $\mathbf{G}$.

A subscript following $\langle \dots \rangle$ indicates the quantity over which the average is taken, e.g. $\langle \mathrm{Y} \rangle _{\mathrm{random \ V}}$ denotes the average of $\mathrm{Y}$ over different random $\mathrm{V}$.

Further information about a variable may be included in a superscript, e.g. $\mathrm{Y}^{\mathrm{meas}}$,$\mathrm{Y}^{\mathrm{pred}}$ and $\mathrm{Y}^{\mathrm{exact}}$ refer to the measurement, prediction and exact value of $\mathrm{Y}$, respectively. Moreover, we use $Y_j^D$ to denote the contribution from segment $D$ to $Y_j$.

\clearpage
\begin{figure*}[h]
	\begin{center}
\includegraphics [width=\textwidth] {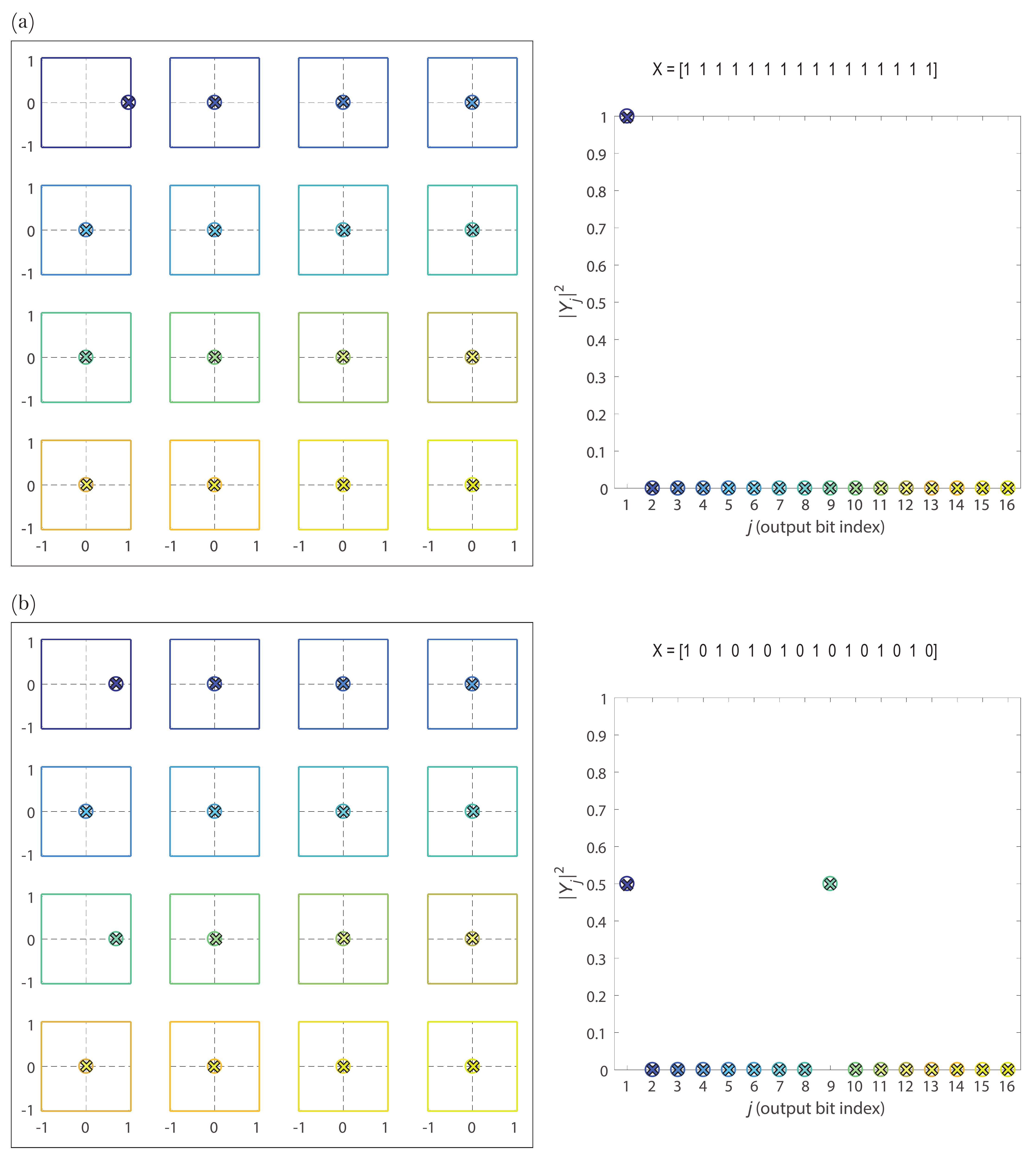}
	\end{center}
\end{figure*}

\begin{figure*}[h]
	\begin{center}
\includegraphics [width=\textwidth] {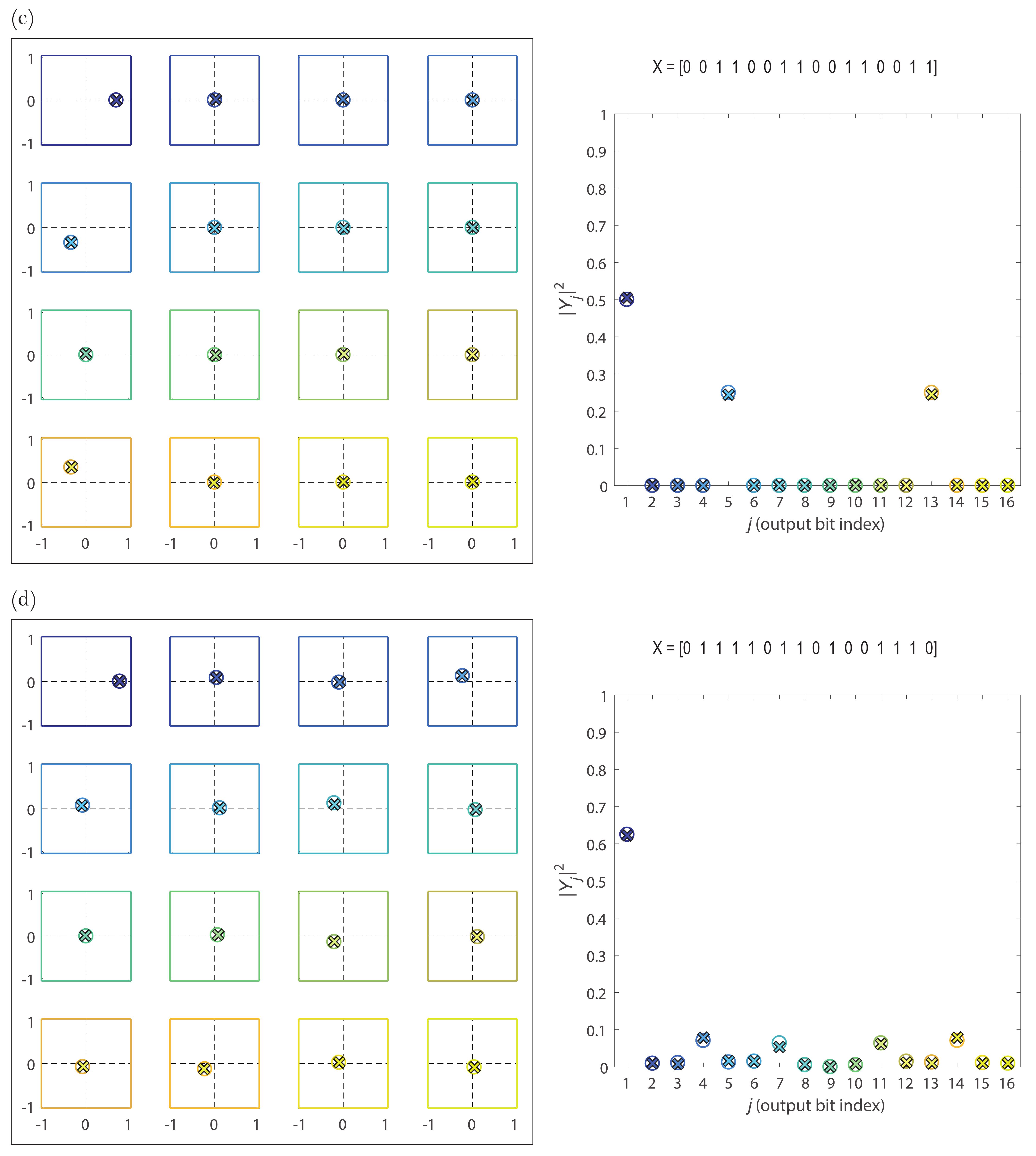}
	\end{center}
\end{figure*}

\begin{figure*}[h]
	\begin{center}
\includegraphics [width=\textwidth] {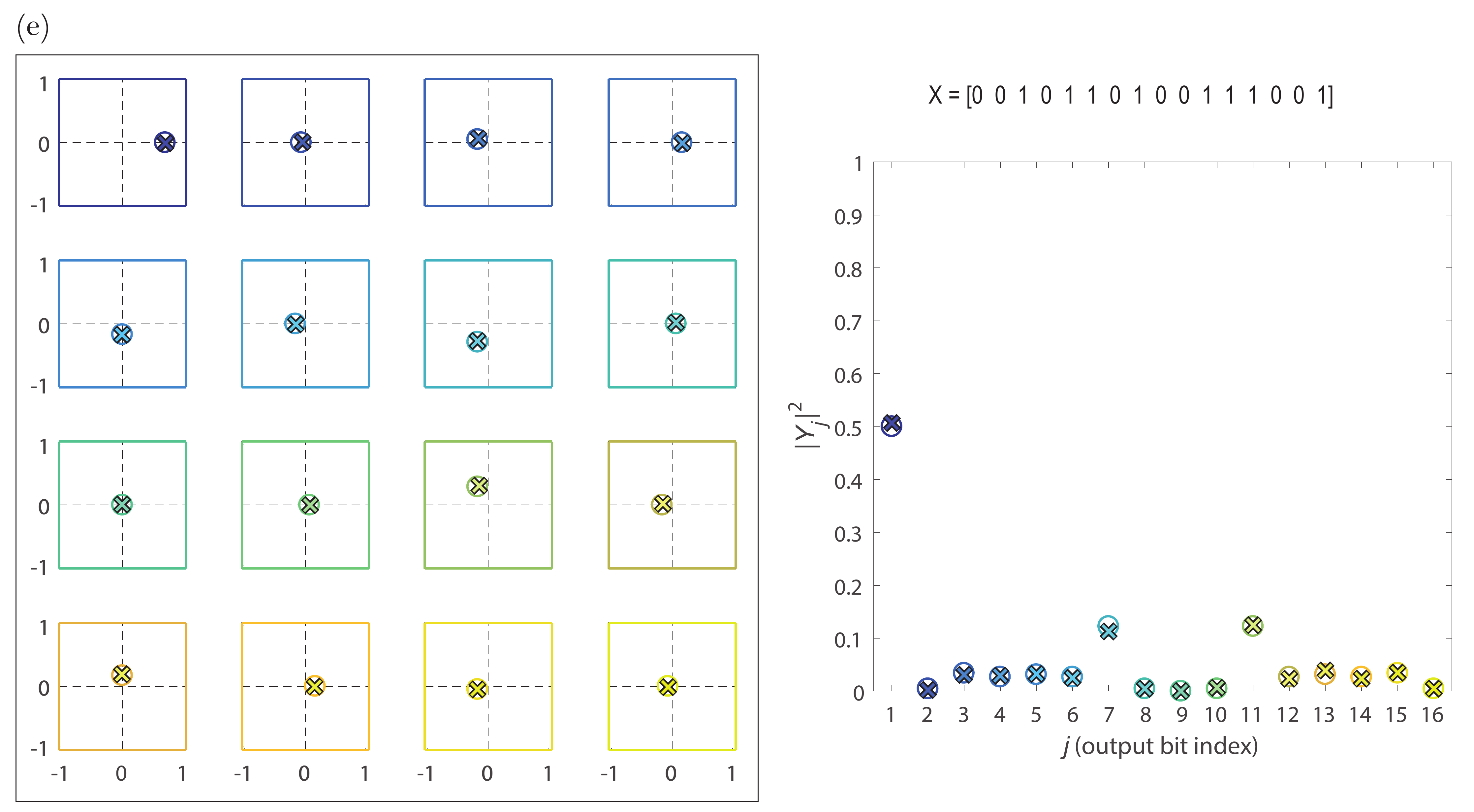}
\caption{Five sample computations with the sequential wave-based analog computation scheme. We show the computation results (crosses) and the exact values (circles) in the complex plane on the left and the corresponding intensities on the right. The input vector $\mathrm{X}$ is indicated for each example.}
	\label{fig_further_examples}
	\end{center}
\end{figure*}

\end{document}